\documentclass[%
 aip,
 amsmath,amssymb,
 reprint,%
]{revtex4-1}

\usepackage{graphicx}
\usepackage{dcolumn}
\usepackage{bm}

\usepackage[utf8]{inputenc}
\usepackage[T1]{fontenc}
\usepackage{mathptmx}
\usepackage{etoolbox}

\makeatletter
\def\@email#1#2{%
 \endgroup
 \patchcmd{\titleblock@produce}
  {\frontmatter@RRAPformat}
  {\frontmatter@RRAPformat{\produce@RRAP{*#1\href{mailto:#2}{#2}}}\frontmatter@RRAPformat}
  {}{}
}%
\makeatother
\begin{document}

\preprint{AIP/123-QED}

\title{Ion-Pairing Enhancement under Osmotic Stress: Disentangling the Effects of Ion and Water Activities}
\title{Ion-Pairing Enhancement under Osmotic Stress: Disentangling the Effects of Ion and Water Activities}

\author{Jay Prakash Singh}
\affiliation{%
Wolfson Department of Chemical Engineering, Technion Israel Institute of Technology, Haifa, Israel%
}

\author{Viatcheslav Freger$^{*}$}
\email{vfreger@technion.ac.il}
\affiliation{%
Wolfson Department of Chemical Engineering, Technion Israel Institute of Technology, Haifa, Israel%
}
\affiliation{%
Grand Technion Energy Program, Technion Israel Institute of Technology, Haifa, Israel%
}
\affiliation{%
Grand Water Research Institute, Technion Israel Institute of Technology, Haifa, Israel%
}

\date{\today}

\begin{abstract}
The dependence of ion pairing on osmotic stress may strongly affect the performance of ionic materials and membranes whose interior is often osmotically stresses, yet quantitative understanding of this dependence and, specifically, the effects of water and ion activities is limited. Motivated by this gap, we analyze the enhancement of ion pairing with osmotic pressure for concentrated aqueous KCl, NaCl, and LiCl solutions using molecular dynamics simulations. Based on rigorous thermodynamic relations, we separate the contributions of ion non-ideality to the pairing constant, varying with osmotic pressure, from other effects including water release and type of ion-pair. Our analysis reveals that ion non-ideality indirectly generates a stronger effect on pairing than the direct one of water release. However, its effect is moderated and may even be reversed for more hydrated pairs by a similarly large and opposite effect of ion-pair non-ideality assigned to varying dielectric properties of the solution and water restructuring upon pairing. The interplay between these contributions, including large and pair type-specific hydration effects on the cost of pairing, explains the observed opposing trends: pairing decreases with osmotic pressure for more hydrated solvent-separated pair types while increasing for contact pairs. The trend becomes more pronounced for more hydrated smaller cations, but was fairly independent of the water model used. The results further suggest that dielectric effects enhanced in ionic materials-and, as a result, larger variations of ion-pair non-ideality, compared with aqueous solutions, should have a more significant impact on pairing than water release.
\end{abstract}

\maketitle

\section{Introduction} 
Addressing the growing demand for efficient ion and salt separations  remains a major challenge in modern technologies\cite{werber2016materials,freger2021polyamide}. The use of ion-selective membranes or nanochannels offers an efficient nature-mimicking solution\cite{fu2025biomimetic,li2025constructing,li2023designing}. However, gaps in understanding of ion transport and thermodynamics in  confined and osmotically stressed environments found in polymers and nanomaterials present a key limitation towards materials development and process design\cite{zhou2020intrapore,geise2014fundamental,epsztein2020towards}. 
Classical mean-field treatments, e.g., of Donnan, Poisson-Boltzmann, or Debye-Hückel (DH), assume full dissociation and capture weak ion-ion interactions \cite{galama2013validity,markovich2014surface,onsager1934surface}. In contrast, ion-pairing comes from strong deviations from mean-field behavior, significantly modifying ion activity and mobility and thus the fixed charge, hydration, reactivity, ion uptake, salt permeability, and conductivity of ionic materials \cite{ben2009beyond,freger2020ion,freger2025ion,oren2024analyzing}. Compared with aqueous solutions, the low permittivity within membranes amplifies ion-ion interactions and thus ion-pairing, posing challenges for modeling   \cite{strathmann2004ion,yaroshchuk2019modelling,kingsbury2020comparison,biesheuvel2022tutorial,oren2024analyzing}.  Following Bjerrum, ion-pairs may be treated \textit{ad hoc} as distinct chemical species in equilibrium with free dissociated ions \cite{bjerrum1968dissociation,robinson2002electrolyte,barthel1998physical,grosberg2002colloquium}. This equilibrium is described by an association constant $K$, measurable in some experiments and readily derived from molecular dynamics (MD) simulation. They typically show a strong concentration dependence, which Bjerrum originally addressed by accounting for free ion non-ideality using the DH theory, to which later studies added semi-empirical corrections, e.g., the Pitzer model\cite{pitzer1986thermodynamics}. 

However, it was also recognized early on that interaction with the solvent (ion hydration) and solvent restructuring around the ions may profoundly affect the values of $K$ \cite{buchner2004complexity,friesen2019cation,freger2020ion}. To this end, ion-pairs were categorized as double solvent-separated (SSIP), single solvent-separated (SIP), and contact ion-pairs (CIP), each with distinct $K$ reflecting progressive shedding and rearrangement of hydration shells upon pairing, counter-balancing closer approach of the two ions and corresponding gain in electrostatic energy \cite{colbin2025ion,roy2017marcus,fennell2009ion,marcus2006ion}. Curiously, the experimental and modeling studies show that $K$ values for CIP and SIP or SSIP pairs vary with salt/ion concentration in a distinctly different and even opposite manner \cite{marcus2006ion}. This drastic difference may not be related to collective ion-ion interactions in the manner of DH theory, yet - to the best of our knowledge - this point was not systematically analyzed. The present study aims to fill this gap and demonstrate how $K$ must respond not only to varying ion activity $a_\pm$ but the solvent (water) activity $a_w$ as well. 

Our motivation to understand the effect of $a_w$, or equivalently osmotic pressure $\Pi = -(RT/V_w) \ln a_w$, where $V_w$ is the partial molar volume of water, stems from the fact that the interior of many charged polymers, membranes, and nanomaterials is osmotically stressed, i.e., $a_w \ll 1$. As a result, direct thermodynamic gain from water release to the external phase upon pairing enhances the propensity of ions to form pairs. The magnitude of this effect may be contextualized by assuming near-ideal behavior of water, $a_w \sim \phi_w$, where $\phi_w$ is the water volume fraction within the material. Expulsion of $\Delta n$ water molecules relased upon pairing  will increase the pairing constant $K$ by a factor of $\phi_w^{-\Delta n}$ \cite{freger2020ion}. In many materials, $\phi_w \sim 0.05$--$0.3$ thus releasing just 2--3 water molecules can boost $K$ by 1--3 orders of magnitude  \cite{freger2020ion}.

However, clarifying  this relation in such materials is not trivial, since these systems are multi-component, which makes ion and water activities independent and complex functions of composition.  The effect of water activity on ion pairing also couples to the similar effect of low-dielectric properties and micro-heterogeneity \cite{freger2025ion}. In addition, the presence of polymer chains introduces excluded volume effects, spatial confinement, and heterogeneous solvation environments. 
These complications make it difficult to disentangle various contributions to pairing and examine their relation to the osmotic stress. However, the situation is greatly simplified in aqueous salt solutions, where presence of just two components, water and salt, rigidly couples water and ion activities through the Gibbs-Duhem relation. This offers a convenient starting point for understanding the relation between osmotic stress and ion-pairing, separating this aspect from the complexities associated with the presence of other polymeric or solid components in above materials. The present paper presents such analysis. 

\section{THEORY}

Since both free hydrated ions and all types of ion-pairs  are essentially ion-water complexes, ion-pairing equilibrium in aqueous solutions can be expressed as
\begin{equation} \label{react}
\text{M}^+(\text{H}_2\text{O})_{n_+}  + \text{A}^- (\text{H}_2\text{O})_{n_-} \rightleftharpoons \text{MA} (\text{H}_2\text{O})_{n_p} + \Delta n \ \text{H}_2\text{O},
\end{equation}
where $n_+$, $n_-$, and $n_p$ are the numbers of water molecules hydrating the cation M$^+$, anion A$^-$, and ion-pair MA, respectively, and $\Delta n = n_+ + n_- - n_p$ is the number of water molecules released upon pairing. For salts of monovalent ions, the corresponding thermodynamic \textit{activity-based} equilibrium constant must be defined as
\begin{equation} \label{Keq}
K_{eq}= \frac{a_p a_w^{\Delta n}}{a_s} 
= \frac{x_p}{x_+ x_-} \frac{\gamma_p}{\gamma_+ \gamma_-} a_w^{\Delta n} 
= K \frac{\gamma_p}{\gamma_{\pm}^2} a_w^{\Delta n},
\end{equation}
where $a_s = a_+ a_-$ and $a_w$ are salt and water activities, $x$'s and $\gamma$'s are respective mole fractions and activity coefficients, and $\gamma_\pm = (\gamma_+ \gamma_-)^{1/2}$. Here, $K = x_p/(x_+ x_-)$ is the \textit{concentration-based} association ``constant" obtained from simulations or experiments, based here on concentrations expressed as mole fractions. Eq.~\ref{Keq} highlights the fact that $K$ deviates from $K_{eq}$ due to non-ideal ion activities and varying water activity. The latter thus affects  $K$ directly through $a_w^{\Delta n}$ and indirectly via intra-pair free-energy change, embedded in $K_{eq}$ and $\gamma_p$ and associated with the electrostatic energy, ion solvation, and water release and restructuring. The indirect effects are the focus of the present study, whose primary goal is to devise the methodology for their quantitative analysis using Eq.~\ref{Keq} as the rigorous thermodynamic basis, complementary to insights into its molecular details or dynamics \cite{gierst1966ion,saveant2001effect,saveant2008evidence,marcus1998ion,hefter2006spectroscopy}. 

To isolate and quantify thermodynamic contributions of different terms in Eq. \ref{Keq} while avoiding the complexities of ternary polymer systems, we investigate here aqueous salt solutions where water and salt activities are rigidly related through the Gibbs-Duhem equation~\cite{atkins2023atkins}, 
$x_w d\ln a_w = - x_s d\ln a_s = - x_s d\ln (\gamma_\pm x_s)^2$,
where $x_s=1-x_w$ is the total salt mole fraction, including free ions and pairs. Another crucial point is that ion-pairing does not involve formation of a chemical bond in the standard sense, whereby the free energy of bond formation $\Delta G = RT \ln K_{eq}$ becomes latent and is not considered when the activity $a_p$ of the newly formed ion-pair species is computed, e.g., as the water activity for the reaction H$^+$ + OH$^- \leftrightarrow$ H$_2$O. Bjerrum's treatment essentially assumes $\gamma_\pm$ and $\gamma_p$ include only ion-ion interaction, totally ignoring hydration and water structure, and includes in $\Delta G$ the electrostatic gain and entropy loss upon pairing, which is equivalent to $K_{eq} = 1 /\gamma_p$ with $\Delta n=0$ and $\gamma_p$ removed from Eq. \ref{Keq}. Yet, when water is added to the picture, it is more convenient to retain ions and water as individual species within ion-pairs and view them simply as distinct states of ions and water, in equilibrium with their free state. Thus, when both $\gamma_\pm$ and $\gamma_p$ explicitly address both ion-ion and ion-water interactions in the free and paired states, the fact they are in equilibrium requires that $\Delta G=0$ and $K_{eq}=1$. With this relation, Eq.~\ref{Keq} simplifies to

\begin{equation} \label{K}
K = \frac{\gamma^2_\pm}{\gamma_p} a_w^{-\Delta n}.
\end{equation}

This relation describes the fundamental connection between the concentration-based pairing constants and the thermodynamic state of water, ions, and ion-pairs in solution. However, in view of the differential nature of the Gibbs-Duhem equation, for analyzing 
the variation of $K$ with concentration, it is more convenient to use it in a form differentiated with respect to the dimensionless osmotic pressure $-\ln a_w = \Pi V_w/RT$. Thus Eq.~\ref{K}, becomes

\begin{equation} \label{K_vs_aw}
-\frac{d \ln K}{d \ln a_w} = - \frac{d \ln \gamma^2_\pm}{d \ln a_w} + \frac{d \ln \gamma_p}{d \ln a_w} + \Delta n =  \left( \frac{1 - x_s}{x_s} + 2\frac{d \ln x_s}{d \ln a_w} \right) + \Delta n^* ,
\end{equation}
where $\Delta n^* = \Delta n  + d \ln \gamma_p / d \ln a_w$ is an effective parameter that conveniently combines  the direct effect of water release and indirect effects of ion-pair non-ideality due to solvation and intra-pair interactions and structure. It may be interpreted as the number of water molecules, whose entropy of release is equivalent to the entropy of actual release plus excess free energy of the ion-pair. As will be seen below, $\Delta n^*$ is derivable from simulated data in a model-independent manner and, when combined with physical constraints, may provide useful insights into the relative magnitudes of the two contributions.

Eq. \eqref{K_vs_aw} is the key relation used below to analyze the MD data on ion-pairing. It highlights the fact that the full effect of varying water activity (osmotic pressure) on pairing is expressed by the derivative $-\frac{d \ln K}{d \ln a_w}$, rather than just water release ($\Delta n$). A similar relation was proposed to describe the effect of water and salt activities on protein-polyelectrolyte complexation in solutions~\cite{record1978thermodynamic, walkowiak2021interaction}. Consistency with the Gibbs-Duhem equation also makes it convenient to define $x_s$ as including all dissolved salt, paired and non-paired, thus $\gamma_\pm$ is defined to match $a_s = (\gamma_\pm x_s)^2$ and all $K$'s are defined for total salt as the reference state. Thus defined $K$'s are also what integration of radial distribution function (RDF) obtained in MD simulations yields (Eq.~\ref{K_RDF} below). However, using full speciation of free ions and pairs deduced from RDF, $K$'s may be readily converted to those redefined using free ions as the more common reference state.\cite{atkins2023atkins}  

In the last expression in Eq. \eqref{K_vs_aw}, the term in brackets expresses the effects of salt non-ideality on observed $K$, varying with salt concentrations and rigidly related to variation of water activity. Notably, this term is directly derivable from the computed or experimental variation of osmotic pressure with salt concentration\cite{hamer1972osmotic}, $a_w$ vs $x_s$. In this manner, $\Delta n^*$ for a specific type of ion-pairs, CIP, SIP and SSIP, could be deduced and analyzed from computed variation of respective $K$.

\section{METHODS: MD SIMULATIONS}
\begin{figure}
    \centering
    \includegraphics[width=1.0\linewidth]{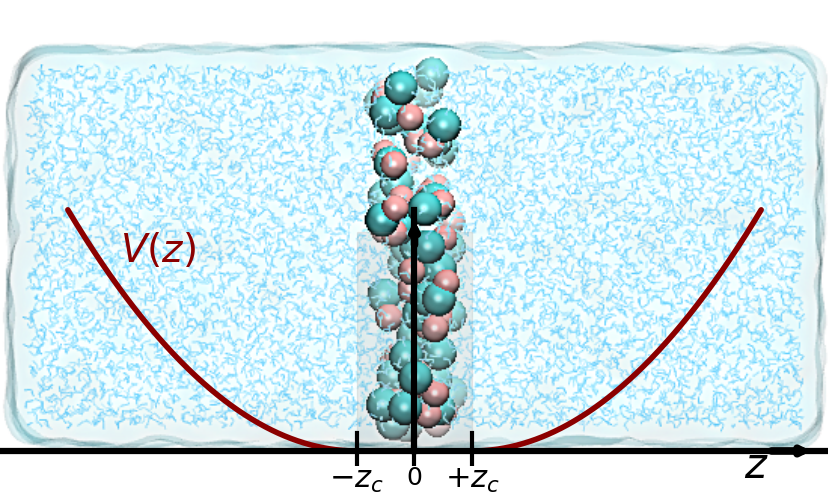}
    
 \caption{
Representative snapshot of water molecules and ions from the molecular dynamics simulation, with the half-harmonic confining potential $V(z)$ (dark red curve) overlaid. The gray shaded rectangle highlights the region of flat potential between $-z_c$ and $+z_c$, where ions are freely diffusing. The potential increases quadratically outside this central region, serving as a confinement mechanism. The vertical arrow denotes the $V(z)$ axis and the horizontal arrow denotes the $z$-axis. The labels $-z_c$, $0$, and $+z_c$ indicate the boundaries and center of the flat potential region.  } 
\label{fig:model}
\end{figure}

To follow consistently the approach outlined above, we used MD to compute \textit{both} the different types of ion-association constants $K$ \textit{and} osmotic pressure $\Pi$, ultimately converted to $a_w$, for monovalent salts $\mathrm{NaCl}$, $\mathrm{KCl}$, and $\mathrm{LiCl}$ over a broad concentration range. The simulation cell initially had dimensions of $\sim 40 \times 40 \times 150~\text{\AA}^3$ and contained $\sim 6500$ water molecules together with $N=50$ cations and $N=50$ anions. Following Luo and Roux~\cite{luo2010simulation}, ions were confined to a bounded region in water, while allowing water exchange with the surrounding bulk. The region was both pressurized and osmotically stressed by subjecting ions to a flat-bottom half-harmonic confining potential along the $z$ axis,
\begin{equation}
\label{Vi}
V(z)=
\begin{cases}
0, & |z|\le z_c,\\
\frac{k}{2}(z-z_c)^2, & |z|>z_c,
\end{cases}
\end{equation}
with spring constant $k=5~\mathrm{kcal\,mol^{-1}\,\AA^{-2}}$. This potential confines ions to $-z_c<z<z_c$ thus the salt concentration and osmotic pressure were increased by reducing $z_c$ and thus compressing region as shown in Fig. \ref{fig:model}.

Simulations were performed using NAMD~\cite{phillips2005scalable} with the CHARMM36 force field. Periodic boundary conditions were applied in all dimensions, and long-range electrostatic interactions were treated using the particle mesh Ewald method~\cite{zheng2017proton}. Water was modeled using the TIP3P model, augmented with the NBFIX corrections of Yoo and Aksimentiev~\cite{yoo2012improved,yoo2018new} and, for some key result, using TIP4P/2005 model. The  TIP3P+NBFIX model was specifically propsoed and parameterized to reproduce experimental osmotic pressures and ion activity data in concentrated salt solutions, which is verified in Sections S2 and S3 of the SM. To examine the dependence of the parameters of interest on the force-field and water model, key simulations for KCl and LiCl were repeated using the TIP4P/2005 water model. This rigid four-site potential, consisting of three fixed charges and one Lennard-Jones center, was parametrized to reproduce the temperature of maximum density, ice polymorph stability, and other key target properties. The model has demonstrated impressive performance in predicting a wide range of properties, including thermodynamic quantities, dielectric constant, pair distribution functions, and diffusion coefficients \cite{abascal2005general}. 

We clarify upfront that, although  TIP3P-NBFIX model correctly captures the osmotic behavior of salt solutions, neither force field is supposed to reproduce the free energy of the solid salts and thus cannot predict salt solubility. The computed osmotic pressures for NaCl and KCl then extend beyond the solubility limit; such nonphysical data points are shown as open symbols in figures and only meant to highlight the overall trends. More importantly, we presume that the  RDFs and paring constant computed by the two models may yield mutually consistent trends and parameters of interest, such as $\Delta n$ and $\Delta n^*$, yet we may not claim quantitative agreement with experimental data, which are still subject to large uncertainties (see below). For this reason, we did not use more advanced models, such as polarizable force-fields, which could improve structural detail but would increase computational cost but were unlikely to change key conclusions.

For each salt, systems were equilibrated in the NPT ensemble, followed by NVT production runs; all reported quantities were evaluated using the equilibrated box dimensions. Since the confined region remains in equilibrium with bulk water, the osmotic pressure equals the mechanical pressure exerted by the confining potential, i.e., $\Pi = P = \langle F \rangle/A$, where $A$ is the instantaneous cross-sectional area of the simulation box, normal to $z$-direction. The mean force acting on the ions was computed as
\begin{equation}
\label{Fi}
\langle F \rangle =
-\frac{1}{2N}
\left\langle
\sum_{i=1}^{N}
\left|
\frac{\partial V_i}{\partial z}
\right|
\right\rangle
=
\frac{k}{2N}
\left\langle
\sum_{i, |z_i|>z_c}^{N_P}
|z_i-z_c|
\right\rangle ,
\end{equation}
where the last sum runs over the $N_P$ ions extending beyond the flat region of the potential, and the factor of $1/2$ accounts for averaging forces from opposite sides. As expected of corrected force-fields~\cite{yoo2012improved, yoo2018new}, thus obtained dependence of osmotic pressure agrees well with experimental data for all analyzed salts (see Section S2 in SM)~\cite{hamer1972osmotic}.

\begin{figure*} [!ht]
    \centering
    \includegraphics[width=1.0\linewidth]{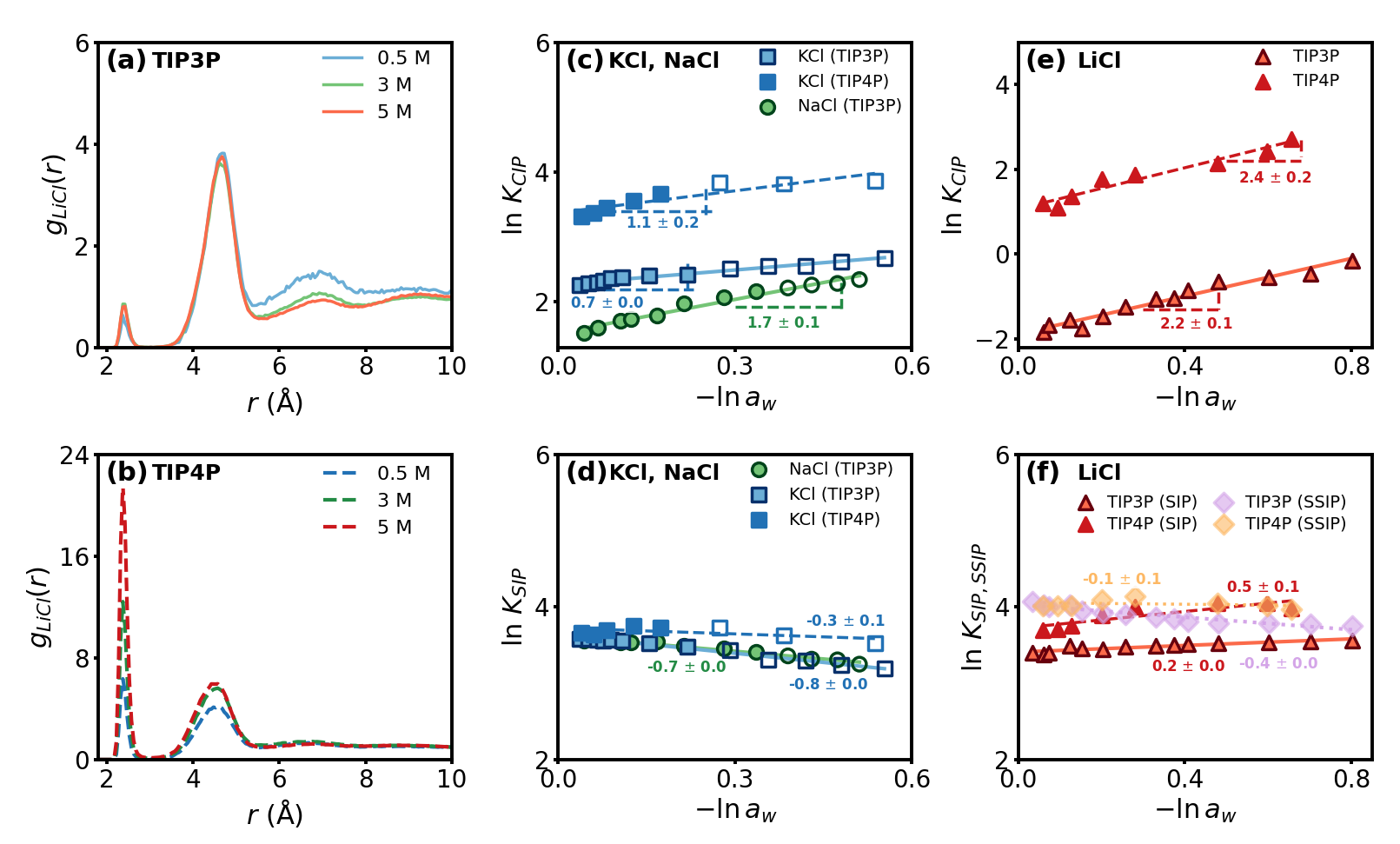}

\caption{
(a,b) Cation-anion radial distribution functions (RDFs), $g_{\mathrm{LiCl}}(r)$, for LiCl at different concentrations using the TIP3P+NBFIX (a) and TIP4P (b) water models. (c--f) Computed association constants as a function of water activity Eq.\ref{K_RDF}, plotted as $\ln K$ versus $-\ln a_w$: $\ln K_{\mathrm{CIP}}$ for KCl and NaCl (c), $\ln K_{\mathrm{SIP}}$ for KCl and NaCl (d), and $\ln K_{\mathrm{CIP}}$ for LiCl (e). (f) $\ln K_{\mathrm{SIP}}$ and $\ln K_{\mathrm{SSIP}}$ for LiCl using TIP3P+NBFIX and TIP4P. The slopes of the linear fits quantify the effective changes in water release, $\Delta n_1^*$ and $\Delta n_2^*$, associated with the corresponding ion-pairing CIP, SIP and SSIP states, with the fitted values indicated in the respective panels. Open symbols indicate simulated
data beyond salt solubility limit.
}
     \label{fig:K_computed}
\end{figure*}

Ion-association constants were extracted from the cation-anion RDFs $g_{\mathrm{M{-}A}}(r)$. Figures \ref{fig:K_computed}(a) and (b) illustrate the computed RDFs for LiCl solutions for the two force-fields, highlighting the differences in the intensity of the first correlation (CIP) peak and the propensity of TIP4P model to exaggerate CIP pairing. Analogous results for KCl and NaCl are shown in Section S1 and Fig. S1 in SM. Based on RDFs, $K_{\text{CIP}}$ and $K_{\text{SIP}}$ were computed using Eq. \ref{K_RDF}
\begin{equation} \label{K_RDF}
K = 4\pi \int_{r_{\min}}^{r_{\max}} g_{\mathrm{M{-}A}}(r)\, r^2\, dr ,
\end{equation}
where the integration limits correspond to RDF minima separating contact ion pairs (CIP), solvent-shared ion pairs (SIP), and more distant free-ion configurations; for LiCl, an additional solvent-separated ion-pair (SSIP) peak is resolved, thus $K_{\text{SSIP}}$ was computed as well. The raw $K$ values (in \AA$^3$) were converted to reciprocal ion mole fraction units by normalization with the total molar content of the confined volume (see Section S1 in Supplementary Material). This yields mole fraction-based association constants suitable for direct comparison with Eq.~\eqref{K_vs_aw}.

The simulated RDFs in Figure ~\ref{fig:K_computed}(a) and Figure S1 for other salts in Supplementary Material reasonably reproduce ones derived from experimental X-ray and neutron scattering data for ionic solutions \cite{mason2019molecular,kohagen2016accounting,harsanyi2012neutron,bouazizi2008structural}. It must be noted that cation-anion RDFs deduced form scattering experiments are prone to significant noise and uncertainties, as they have to be separated from strong background scattering, where cation-anion correlation peaks also overlap with far more abundant correlations, e.g., cation-oxygen. Nevertheless, there is much similarity of the present RDFs and their variation with concentration and experimental reports, especially, for KCl~\cite{mason2019molecular} and NaCl~\cite{kohagen2016accounting,bouazizi2008structural}. Some discrepancies may be noted for LiCl, e.g., even for TIP3P-NBFIX model, the CIP peak seems to be more pronounced in the spectra derived from scattering experiments, however, their noise and uncertainties are also substantial~\cite{harsanyi2012neutron}.

\section*{RESULTS AND DISCUSSION}
\subsection{Concentration-dependence of pairing constants }

Figures \ref{fig:K_computed}(c) to (f) plot the variation of computed association constants with  $-\ln a_w$. Notably, $K_{\text{SIP}}$ for KCl and NaCl and $K_{\text{SSIP}}$ for LiCl display near-zero or even slightly negative slopes, i.e., decrease with salt concentration, while $K_{\text{CIP}}$ shows an opposite trend. We will see below that the decreasing trend of more hydrated pairs is non-trivial to explain, yet indications of similar trends of CIP and SIP pairs were reported in scattering studies~\cite{bouazizi2006local}. The trends of $K$ increasing with salt concentration for less hydrated pairs and decreasing for more hydrated pairs also agree with the dielectric relaxation spectroscopy (DRS) data by Buchner et al. for MgSO$_4$ and MgCl$_2$ salts that pair far more strongly thus DRS may differentiate between pair types \cite{buchner2004complexity, friesen2019cation, hefter2020dielectric}. The decreasing water activity $a_w$ should indeed favor less hydrated ion-pair species with larger $\Delta n$, i.e., suppress  more or promote less SIPs compared with CIPs (Eq. \eqref{K_vs_aw}). Yet, in addition to this direct effect, each individual trend may also reflect the indirect contribution of varying ion and ion-pair non-ideality (activity coefficients) analyzed below. 

Ion activity coefficients in salt solutions commonly display a non-monotonic trend, first decreasing with concentration in dilute solutions, in agreement with the DH theory, and thereafter increasing  \cite{pitzer1986thermodynamics}, which was recently attributed to decreasing polarizability of solutions \cite{shilov2015role,valisko2017activity} (see Section S3 in SM). As the present concentrations are all large, $\gamma^2_\pm$ increases, which should promote pairing, but, as follows from Eq. \eqref{K_vs_aw}, the effect may be offset by $\gamma_p$. Below we analyze this point more quantitatively, by evaluating $\Delta n^*$ associated with formation of each type of ion-pairs, as well as transition between them (see also Section S4 in SM).  

\subsection{Separating the effects of $\gamma_\pm$ and $a_w$: pair-to-pair transitions}
\begin{figure}
    \centering
    \includegraphics[width=1.0\linewidth]{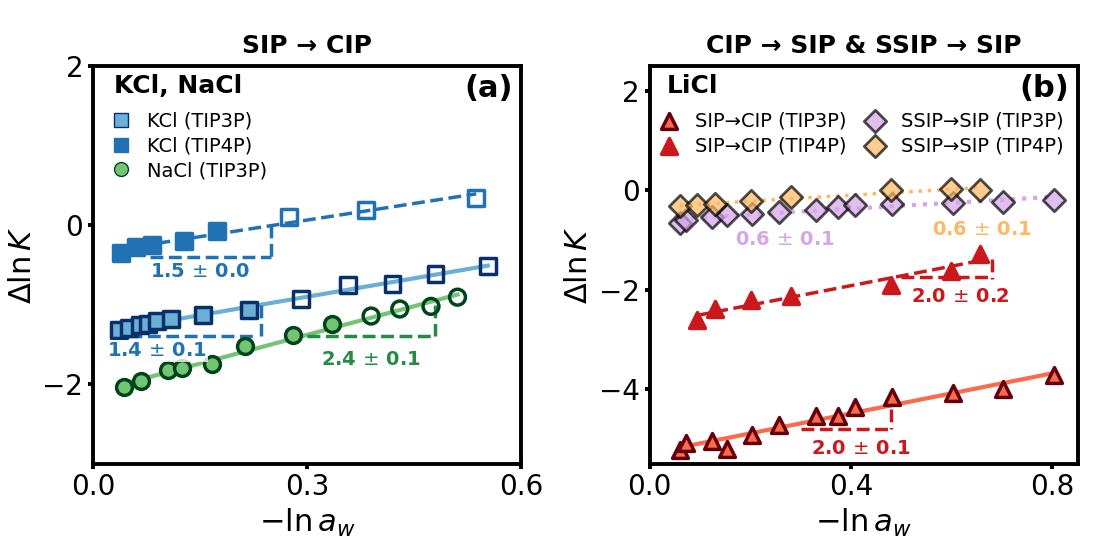}

\caption{
(a) $\Delta \ln K$ for the SIP $\rightarrow$ CIP transition in KCl and NaCl using TIP3P+NBFIX and TIP4P water models. (b) $\Delta \ln K$ for the SIP $\rightarrow$ CIP and SSIP $\rightarrow$ SIP transitions in LiCl. The slopes of the linear fits provide the effective water release, $\Delta n_1^*$ and $\Delta n_2^*$, for the corresponding SIP to CIP and, SSIP to SIP transitions, with the fitted values indicated in the plots. Open symbols indicate simulated
data beyond salt solubility limit.
}
     \label{fig:ionc_removed}
\end{figure}

We first note the ion-activity term must be identical for all pairs, therefore it may be canceled out by analyzing  $\Delta \ln K$ of \textit{transition} between them, i.e., $\ln (K_{\text{CIP}}/K_{\text{SIP}})$ for all three salts and $\ln (K_{\text{SIP}}/K_{\text{SSIP}})$ for LiCl. Figure~\ref{fig:ionc_removed} displays the computed variation of $\Delta \ln K$ versus $-\ln a_w$, whose slopes are identical with the differences in respective $\Delta n^*$. They are denoted as $\Delta n^*_1$ and $\Delta n^*_2$ for the transitions SIP  →  CIP and SSIP  →  SIP, respectively. These numbers stay positive throughout, as expected of progressive water release.
Curiously, both $\Delta n^*_1$ and $\Delta n^*_2$ remain fairly constant with varying osmotic pressure, i.e., the trends of $\Delta \ln K$ versus $-\ln a_w$ are close to linear. Furthermore, average $\Delta n^*_1$ for KCl, NaCl, and LiCl, equal to 1.5, 2.0, and 2.5, respectively, are all close to about two water molecules and surprisingly similar, despite much different hydration numbers and hydration energies of respective cations \cite{marcus2007solvent}.

Although the absolute values of $\Delta \ln K$ obtained using TIP3P+NBFIX and TIP4P force fields largely differ, especially for CIP pairs, the $\Delta n^*_1$ and $\Delta n^*_2$ values derived form their variation with concentration do not differ, considering the error of the fitted slopes. We presume that the numbers $\Delta n^*_1 \sim 1.4-1.5$ for KCl, $\Delta n^*_1 \sim 2.4$ for NaCl, and $\Delta n^*_1 \sim 2$ and $\Delta n^*_2 \sim 0.6$ for LiCl may reflect the genuine number of water-molecules released upon respective transitions. Their model independence suggests that, for each specific ions and pair-to-pair transitions, they are mainly dictated by the steric (entropic) factors or packing and release of water molecules regardless of the (force-field dependent) ion-ion and water-ion interaction. Specifically, loosely bound water is rearranged and released in about the same manner regardless of the specific force-field, which thus mainly affects the absolute value of $\Delta \ln K$ and not its variation with water activity, i.e., the trends may shift vertically but keep the same slope. Conversely, tightly bound water does not rearrange and thus affects neither the absolute value of $\Delta \ln K$ nor the slope.

It is also notable that the small $\Delta n^*_2 < 1$ for SSIP  →  SIP transition of LiCl somewhat deviates from the standard view of the process SSIP → SIP → CIP whereby each transition releases commensurate number of water molecules \cite{marcus2006ion,marcus2005electrostriction}. However, its values and differences in $\Delta n^*$ may also reflect and even be dominated by changing ion and ion-pair activities, as discussed below.
The transitions between ion-pairs provide some insights but, with ion activity canceled out, $\Delta n^*_1$ and $\Delta n^*_2$ may substantially differ from respective $\Delta n^*$ for pairing of \textit{free ions}. The latter may be quantified by explicitly eliminating the contribution of the ion-activity term in brackets in Eq. \eqref{K_vs_aw} from the computed $K$, as discussed next.

\begin{equation}
\label{eq:BET}
q =\frac{1- x_s}{x_s} = H \frac{C a_w}{(1 - a_w) [1 + (C - 1)a_w]},    
\end{equation}
where $q$ is the number of water molecules per salt. The BET model approximates collective interaction of all water molecules present in solution with salt as a multilayer adsorption, where the salt (cation + anion) acts as a primary adsorption site, which was found to describe well water activity in concentrated electrolyte solutions \cite{robinson2002electrolyte}.
\subsection{Separating the effects of $\gamma_\pm$ and $a_w$: ion-to-pair transitions}
This ion activity term may be obtained directly and integrated using the dependence of $x_s$ on $a_w$ or osmotic pressure, as another outcome of the present simulations. Although experimental osmotic and ion-activity data are available for all salts, the use of simulated osmotic data was preferred here to keep consistency with the pairing-constant results, for which the experimental data are scarce and highly uncertain. 
To facilitate the analysis and also minimize scatter and numerical artifacts, this dependence was fitted to Brunaer-Emmet-Teller (BET) isotherm,

\begin{figure*}
    \centering
    \includegraphics[width=1.0\linewidth]{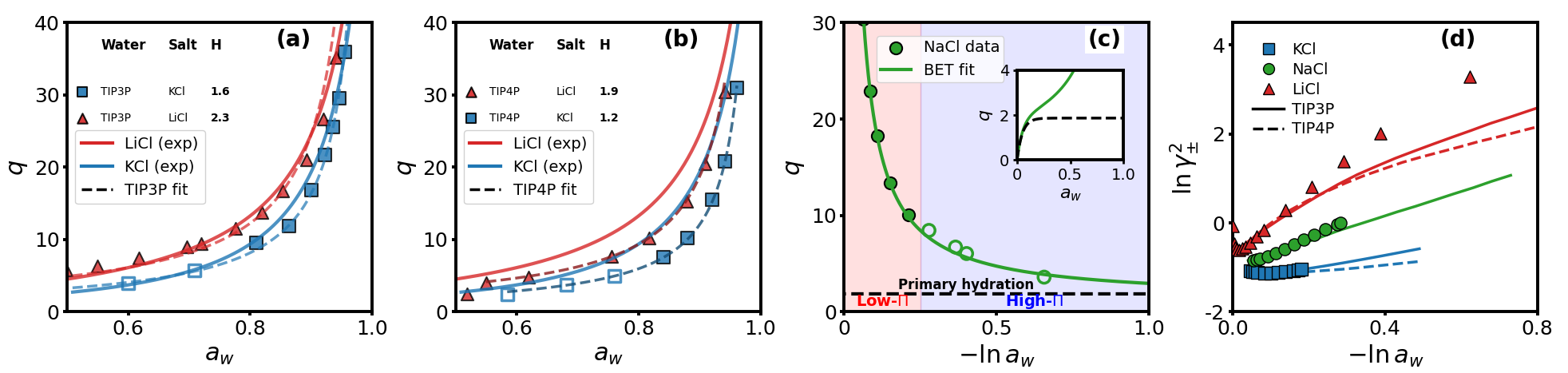}
    \caption{
Variation of the number of water molecules per salt $q$ with water activity, $a_w$. (a) Computed solution composition as $q$ vs. $a_w$ for KCl, NaCl, and LiCl (symbols) and corresponding fits to the BET isotherm, Eq. \eqref{eq:BET} (dashed lines) for TIP3P+NBFIX model; legend indicates corresponding fitted $H$ (hydration numbers); solid lines are BET fits to experimental data. (b) same as (a) for TIP4P model.  (c) NaCl data illustrating the use of BET fit and Langmuir isotherm (primary hydration)  with the same $H$ and $C$ values for differentiating between the osmotically stressed ($q \sim H$, high-$\Pi$) and dilute ($q \gg H$, low-$\Pi$) regimes. (d) Dependence of the ion activity term $\ln \gamma^2_{\pm}$ on the negative logarithm of water activity $-\ln a_w$. Experimental data (symbols) were taken from Ref.\cite{hamer1972osmotic}. Solid and dashed lines are derived from fits to Eq. \ref{eq:BET} as described in SI for results obtained using TIP3P+NBFIX and TIP4P models, respectively. For KCl and NaCl data points are limited by solubility, but fits extrapolate to include all simulated points.
}
    \label{fig:osm_bet}
\end{figure*} 
Figure  \ref{fig:osm_bet} (a) shows the BET model produces reasonable fits to the  data simulated using TIP3P-NBFIX model as well as experimental data (all data may be found in Section S2 in SM ), with a close agreement between the two. On the other hand, TIP4P model results in Figure  \ref{fig:osm_bet} (b) fit the BET model as well but show significant deviations from experimental values. The parameters $H$ and $C$ of the BET model are interpreted, respectively,  as the number of water molecules directly interacting with salt ions (primary hydration shell) and the strength of this interaction in excess of water-water interaction, respectively. Fitted $H$ then supplies a purely thermodynamic estimate of the hydration number, which is a fairly obscure quantity dependent on the method of measurements \cite{marcus2007solvent,marcus2006ion}. The fitted $H$ values are indicated in Figure~\ref{fig:osm_bet} and in Figure S2 in SI and follow the established trend of cation hydration  $\mathrm{K^+ < Na^+ < Li^+}$. However, they are all confined to the range between 1.6 and 2.6 and thus are smaller and much closer to each other than other commonly considered mobility-based hydration numbers\cite{robinson2002electrolyte, marcus2007solvent}. On the other hand, they are commensurate with thermodynamics-based numbers, e.g., as derived from electrostriction \cite{marcus2005electrostriction}. Apparently, they reflect the strongest direct-contact binding and thus may be compared with the number $\Delta n^*$ of equivalent released water molecules.

In this context, as illustrated for NaCl in Figure~\ref{fig:osm_bet}(c),  it is also expedient to differentiate qualitatively between the high-$\Pi$ (osmotically stressed) and low-$\Pi$ range of $a_w$. In the former, the number of water molecules per salt is of the order of $H$ thus their release or rearrangement is costly. In contrast, in the low-$\Pi$ range, above $a_w \sim 0.75$, water is relatively abundant, and it is released and rearranged more readily. Below we focus on the the latter range, as the high-stress range is physically realizable only for LiCl and goes beyond the solubility of KCl and NaCl. 

\begin{figure} [!htb]
    \includegraphics[width=1.0\linewidth]{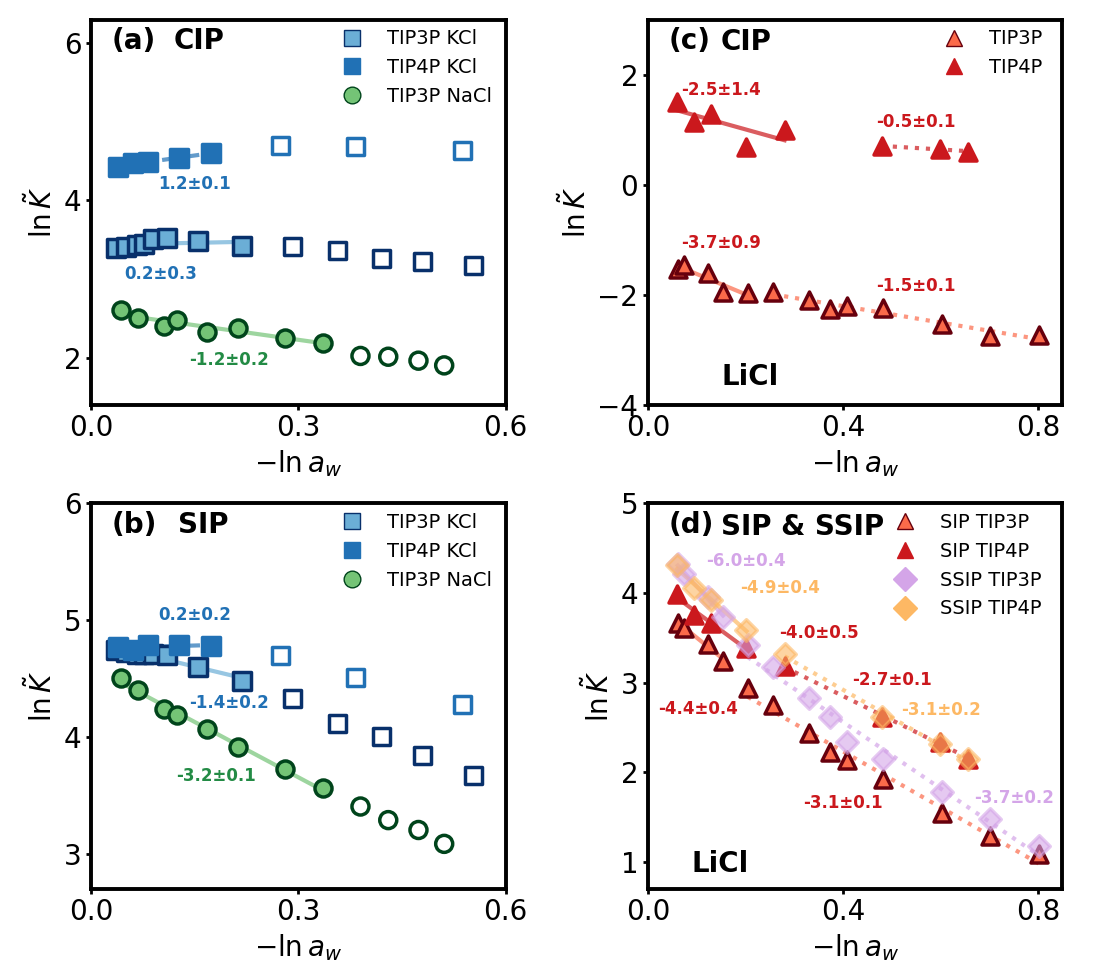}
     \caption{
The variation of $\ln \widetilde{K} = \ln K - \ln \gamma^2_{\pm}$  versus $-\ln \mathit{a_w}$ for CIP (a,c), and SIP and SSIP (b,d) pairs of KCl and NaCl (a,b), and LiCl (c,d) computed using TIP3P+NBFIX and TIP4P models. The correction term computation based on BET fit to osmotic data is as described in SI. The slopes of linear fit to each trend, corresponding to $\Delta n^*$, in low- and high osmotic pressure regions are indicated.  Open symbols indicate simulated data beyond salt solubility limit.  
}
    \label{fig:K_ion_removed}
\end{figure}
The BET fits of simulated data were used to calculate the ion-activity contribution - the term in brackets in Eq. \ref{K_vs_aw}, - as elaborated in Section S4 in the SM. Figure~\ref{fig:osm_bet}(d) shows the obtained variation of computed $\ln \gamma^2_{\pm}$ with osmotic pressure, which reasonably agrees with experimental ion activities~\cite{hamer1972osmotic} for TIP3P+NBFIX model but somewhat deviates for TIP4P (see also Section S3 in SM). When subtracted from the simulated $K$ values, it produces variation of "corrected" constants  $\ln \widetilde K = \ln K - \ln \gamma^2_{\pm}$, in which the ion-activity contribution is eliminated. Figure \ref{fig:K_ion_removed} presents the resulting dependence of $\widetilde K_{\text{CIP}}$ and $\widetilde K_{\text{SIP}}$ on the dimensionless osmotic pressure $-\ln a_w$ for the three salts, as well as $\widetilde K_{\text{SSIP}}$ for LiCl. The slope of $\ln \widetilde K$ versus $-\ln a_w$  then directly yields $\Delta n^*$ (cf. Eq.~\ref{K_vs_aw}).

\subsection{Discussion: the trends of $\ln \widetilde K$ versus $\ln K$ and $\Delta \ln K$}
We note that the trends of $\ln \widetilde K$  in Figure \ref{fig:K_ion_removed} show differences between cations and ion-pair types. For pairs of more hydrated NaCl and LiCl salts, the trends display a pronounced decrease with osmotic pressure, i.e., negaitve $\Delta n^*$ values, whose absolute value is larger for SIP and SSIP pairs and for LiCl salts. 
On the other hand, in the phsyical range below solubility, KCl displays small $\Delta n^*$ values slopes that are somewhat larger for CIP pairs and TIP3P+NBFIX model and drop to near- or sub-zero values for SIP pairs and TIP4P model. Eventually, when extrapolated to non-physical range above solubility, the slopes turn negative for KCl as well. The different behavior of KCl mainly reflects the weaker variation of its $\ln \gamma_\pm$ with concentration, cf. Fig. \ref{fig:osm_bet}(d).   

Focusing first on the large negative $\Delta n^*$ of more hydrated salts, we note it contrasts the uncorrected constants $\ln K$ in Figure~\ref{fig:K_computed} (c), (d), and (f), displaying much smaller slightly positive or negative slopes for respective pairs. This highlights the fact that the removed ion-activity contribution, promoting pairing, is large and \textit{positive}, while the remaining combined contributions of water release ($\Delta n)$ and ion-pair activity is similarly large and \textit{negative}. Since $\Delta n$ can not be negative by definition and is supposed to promote pairing as well, the water release presumed in Eq. \ref{react} is apparently largely outweighed by the ion-pair activity variation, $d\ln \gamma_p/d\ln a_w$. The latter derivative is therefore substantial and negative, similar to $d\ln \gamma^2_\pm/d\ln a_w$ of the ions. This conclusion does not contradict the positive slopes of $\Delta n^*_1$ and $\Delta n^*_2$ in Figure~\ref{fig:ionc_removed}. Although comparison is between different pairs rather than pairs and free ions similarly removes (cancels out) the ion activity contribution, a major portion of the negative $d\ln \gamma_p/d\ln a_w$ is canceled out as well, which gives more weight to $\Delta n$. 

Substantial contribution of ion-pair activity term, $d\ln \gamma_p/d\ln a_w$, stands in stark contrast with the original Bjerrum's picture that assigned all non-ideality to free ions, ignoring that of ion-pairs. However, correlated and commensurate variation of $d\ln \gamma_p$ and $d\ln \gamma^2_\pm$ is readily rationalized by adding varying ion hydration to his picture, which may be approximately treated using the primitive model (PM) viewing the solvent as a dielectric continuum. 

Indeed, the Born model expresses the hydration (self-energy) contribution to the activity of the free cation and anion as $\ln \gamma^2_\pm \simeq \lambda (b_+^{-1}+b_-^{-1})$, where $b_\pm$ are the ion diameters and  $\lambda$ is the Bjerrum length, inversely related to the dielectric permittivity of the solution~\cite{freger2020ion}. The approximately linear increase of simulated $\ln \gamma^2_\pm$ with osmotic pressure $-\ln a_w$ in concentrated solutions was previously assigned to decrease of dielectric permittivity with concentration~\cite{shilov2015role,  valisko2017activity}. This decrease is known to occur due to strong immobilization of water within primary hydration shells as well as disruption of hydrogen-bonding network of more distant water molecules, which suppresses dipolar correlations \cite{zhang_klein_2023}. $\lambda$ increasing about linearly with the osmotic pressure is then the key factor controlling variation of $\ln \gamma^2_\pm$ in the examined range. 
Within PM, the difference between $\ln \gamma_p$ and $\ln \gamma^2_\pm$ is the gain in electrostatic energy, $-\lambda/b_p$ upon pairing, where $b_p$ is the distance between the charges or the size of the pair, partly offset by the loss of translational entropy. This yields the classical Bjerrum expression of ion-paring constant $\ln K \simeq const + \ln(b_p/\lambda)+\lambda/b_p$~\cite{freger2020ion}. Given the logarithmic (entropy) term is weakly changing and $b_p \geq (b_+ + b_-)/2$, the difference $\ln (\gamma_p/\gamma^2_\pm)$ is a fraction of $\ln \gamma_p$ or $\ln \gamma^2_\pm$ and all three should vary in a correlated manner. Thus, due to increasing $\lambda$, $d\ln (\gamma_p/\gamma^2_\pm)/d \ln a_w$ should be negative and, according to Eq.~\ref{K_vs_aw}, promote more pairing at larger osmotic pressures and increase $K$, adding to the effect of $\Delta n>0$. This also anticipates that the increase of $K$ with osmotic pressure should become stronger in the order SSIP → SIP → CIP, since $b_p$ gets smaller and  $\Delta n$ larger.  

When these PM predictions are compared with the results in Figures~\ref{fig:K_computed}, \ref{fig:ionc_removed}, and \ref{fig:K_ion_removed}, osmotic pressure indeed promotes pairing more (or suppresses less) in the above order. However, only CIP constants in Figures~\ref{fig:K_computed}(c) and (e) show a pronounced increase. Even in these cases, $\Delta n$ of at least 2 is anticipated for CIP, while the observed slopes are lower or barely exceed 2. In fact, the discrepancy is even larger, given negative $d\ln (\gamma_p/ \gamma^2_\pm)/d\ln a_w$ should increase the slope beyond 2, however, the results indicate its effect is small or, possibly, even negative.

This conclusion is even stronger for SIP amd SSIP pairs. Here, Figures~\ref{fig:K_ion_removed}(b) and (d) display still more negative slopes ($\Delta n^*$ ), indicative of strong increase of $\ln\gamma^2_\pm$ with osmotic pressure, directly seen in Figur~\ref{fig:osm_bet}(d). This anticipates a substantial and positive $d\ln (\gamma_p/ \gamma^2_\pm)/d\ln a_w$ and $K$ increasing with osmotic pressure. Yet Figures~\ref{fig:K_computed} (d) and (f) show very small slopes, slightly positive or negative. This indicates a positive $d\ln (\gamma_p/ \gamma^2_\pm)/d\ln a_w$ that subtracts from and entirely cancels out the pairing-promoting effect of water release ($\Delta n$) and shows no correlation with ion-activity variation $d\ln\gamma^2_\pm/d\ln a_w$, contrary to PM predictions.

It was argued that deviations from the Born model for individual ions are usually not excessive and may be adequately addressed by slight adjustments of ion radii for ions considered here~\cite{babu1999new, duignan2013continuum, shilov2015role,  valisko2017activity, freger2023dielectric}. We then speculate that PM fails most significantly for ion-pair activity, i.e., underpredicts $d\ln \gamma_p/d \ln a_w$ hence also $d\ln (\gamma_p/ \gamma^2_\pm)/d\ln a_w$. Likely, the failure originates from \textit{a major reorganization} of water molecules upon pairing that PM is unable to account for. We hypothesize that, while some water may be released and thus increase entropy, part of the electrostatic gain $-\lambda/b_p$ is expended on \textit{decreasing the entropy} by re-packing water around the ion-pair in a denser and more immobilized manner. Importantly, this The resulting loss of entropy may reverse the gain of water release and offset the electrostatic gain well beyond the simple loss of translational entropy of the ions and substantially modify $\ln \gamma_p$ and its vairiation.  This may be viewed as a specific case of the well-known enthalpy-entropy compensation (EEC) \cite{searle1995application,fox2018molecular,ryde2014fundamental} . 

The present findings and trends seem to agree well with hypothesis. Indeed, water becomes more difficult to rearrange thus the anticipated increase of $K$ with osmotic pressure is more pronounced when

(a) the cation gets smaller, more hydrated, and more kosmotropic thus water structure is denser and less compressible;

(b) the pair type gets tighter, thus rearrangement needs to involve more strongly bound water;

(c) the MD force field tends to over-predict cation-anion attraction and degree of pairing.

We stress that the considered effect refers only to relative changes with osmotic pressure and not absolute values of different pairing constants. The results in Fig. \ref{fig:K_computed} indeed demonstrate that the enhancement of pairing with osmotic pressure increases in the order (a) KCl < NaCl < LiCl, (b) SSIP < SIP < CIP, and (c) TI3P+NBFIX $\leq$ TIP4P model, in agreement with above predictions. In this order, the relative role of entropy loss decreases, while that of electrostatic gain increases and may deviate less from PM predictions. 

\subsection{Implications for pairing in solutions and ion-selective materials}
The agreement of the trends of pairing constants with the enthalpy-entropy compensation mechanism suggests that the notion of "water release" upon pairing may be fairly misleading. The present results and the EEC logic suggest that the overall effect of water restructuring on entropy in response to electrostatic attraction of ions (enthalpy) is in fact opposite and is more akin to "water capture". The electrostatic gain increases in more concentrated systems due to decreasing polarizability and thus promotes more pairing. It may be attenuated in the manner of EEC by water rearrangement, especially, for looser types of pairs, for which transitioning to a tighter structure produces a larger entropic compensation. However, this compensation may have limits dictated by densest water packing and immobilization and it may become less substantial as electrostatics gets stronger and pairs get tighter.   

Returning now to the possible enhancement of pairing within ionic polymers or other solid ionic materials that motivated the present study, we note that the intra-polymer environment is not only osmotically stressed but also inherently low-dielectric and heterogeneous. The present results suggest that, if the effect of osmotic pressure on pairing involves release  of 1-2 molecules, it should be fairly moderate, even if not negligible. However, a recent study concludes that an order-of-magnitude lower dielectric constants of polymers (hence much larger $\lambda$) may strongly enhance pairing. Given the ions are charged and pairs are neutral, compared with single hydrated ions separated in different pores, formation of a hydrated pair in a water-filled nanopore surrounded by a low-dielectric greatly reduces the dielectric polarization energy and generates an electrostatic gain far exceeding that of pairing in solutions ~\cite{freger2023dielectric}. Thereby, the nature of such enhancement may be reminiscent of the effect of moderately increasing $\lambda$ in concentration solution analyze here ~\cite{freger2025ion}, but its magnitude may be far larger. This ``pore model" was found to describe well ion uptake in neutral polymers~\cite{bannon2025influence}. However, for such a large gain, the compensating entropy loss is likely to reach a limit and become relatively small and insignificant. 

\section{CONCLUSIONS}

In summary, motivated by the need to understand ion pairing in osmotically stressed environments within ionic materials, we analyzed its variation in solutions for three common monovalent salts using MD simulations. Starting from rigorous thermodynamic relations, the direct effect of water activity through water possible release upon pairing ($\Delta n$ of a few water molecules) was separated from that of ion non-ideality (captured by $\gamma^2_\pm$). The results suggest that the magnitude of ion non-ideality contribution to pairing is more significant than that of water release; however, it was largely offset and may even be outweighed by a commensurate effect of  ion-pair activity ($\gamma_p$). While the contribution of $\gamma^2_\pm$ increasing with osmotic pressure is consistent with the primitive model (PM) of salt solution and solution polarizability decreasing with osmotic pressure, the combined effect of $\gamma^2_\pm$ and $\gamma_p$ could be opposite, depedning on the pair-type and contrary to PM predictions. As a result, the trend of pairing increasing with osmotic pressure, as expected form PM and observed for CIP pairs, could be reversed for more hydrated SIP and SSIP pairs. Such behavior was ascribed to short-range solvation and water re-structuring (equivalent to "water capture" rather than release) that occurs more readily around solvent-separated pairs and less hydrated ions, as a special case of entropy-enthalpy compensation.

While the present study is limited to bulk aqueous solutions and does not directly model the complex environment with ionic polymer, materials, or membranes, the thermodynamic framework and the quantitative relationships between pairing constants and thermodynamic parameters such as $a_w$, $\gamma^2_\pm$, $\gamma_p$, and $\Delta n$ established here provide a usefulreference for understanding ion pairing under osmotic stress. The results suggest that dielectric polarization effects, which are enhanced in ionic materials and low-dielectric polymer environments, should have a more significant impact on pairing than water release and restructuring that can be  limited in magnitude. More research is clearly required to systematically evaluate these effects in complex heterogeneous systems, and the baseline established here may serve as a useful starting point for future investigations.

\section*{SUPPLEMENTARY MATERIAL}
The supplementary material provides computational and validation details supporting the main findings. It presents the computed cation-anion radial distribution functions (RDFs) for KCl, NaCl, and LiCl, along with the integration method used to derive concentration-based association constants for contact (CIP), solvent-shared (SIP), and solvent-separated (SSIP) ion pairs. The force-field parameters are validated by comparing simulated water activities and ion activity coefficients against experimental data, with BET isotherm fits showing reasonable agreement for all three salts. Finally, the theoretical framework for correcting the association constants by removing ion-activity effects is derived, enabling the isolation of the water activity contribution ($\Delta n^*$) discussed in the main text.
\section*{SUPPLEMENTARY MATERIAL}

\begin{acknowledgments}
The authors thank the funding sources. V.F. acknowledges the financial support from the Israel Science Foundation (grant 486/22), the Israeli National Institute for Energy Storage (INIES), the Israel Fuel Cells Consortium (IFCC), the Ministry of National Infrastructure, Energy, and Water Resources of Israel (grant no. 222-11-064), the Israel Ministry of Innovation, Science and Technology (grant no. 0005658 within the China-Israel Scientific Research Program), and the Grand Technion Energy Program (GTEP). J.P.S. acknowledges the Lady Davis Foundation Fellowship.
\end{acknowledgments}

\section*{author declaration}
\subsection*{Conflict of interest}
The authors have no conflicts to disclose.

\subsection*{Author Contributions}
\textbf{JP Singh:} Data curation, formal analysis, investigation, methodology, writing - original draft.
\textbf{V. Freger:} Conceptualization, methodology, resources, supervision, project administration, writing - review \& editing.

\section*{Data Availability Statement}
The data that support the findings of this study are available
from the corresponding author upon reasonable request.
\appendix
\renewcommand\refname{ references}

\bibliography{aipsamp}

@article{walkowiak2021interaction,
  title={Interaction of polyelectrolytes with proteins: quantifying the role of water},
  author={Walkowiak, Jacek J and Ballauff, Matthias},
  journal={Advanced Science},
  volume={8},
  number={12},
  pages={2100661},
  year={2021},
  publisher={Wiley Online Library}
}

@article{record1978thermodynamic,
  title={Thermodynamic analysis of ion effects on the binding and conformational equilibria of proteins and nucleic acids: the roles of ion association or release, screening, and ion effects on water activity},
  author={Record Jr, M Thomas and Anderson, Charles F and Lohman, Timothy M},
  journal={Quarterly reviews of biophysics},
  volume={11},
  number={2},
  pages={103--178},
  year={1978},
  publisher={Cambridge University Press}
}

@article{friesen2019cation,
  title={Cation hydration and ion pairing in aqueous solutions of MgCl2 and CaCl2},
  author={Friesen, Sergej and Hefter, Glenn and Buchner, Richard},
  journal={The Journal of Physical Chemistry B},
  volume={123},
  number={4},
  pages={891--900},
  year={2019},
  publisher={ACS Publications}
}

@article{buchner2004complexity,
  title={Complexity in “simple” electrolyte solutions: Ion pairing in MgSO4 (aq)},
  author={Buchner, Richard and Chen, Ting and Hefter, Glenn},
  journal={The Journal of Physical Chemistry B},
  volume={108},
  number={7},
  pages={2365--2375},
  year={2004},
  publisher={ACS Publications}
}

@article{zhang_klein_2023,
  title = {Why Dissolving Salt in Water Decreases Its Dielectric Permittivity},
  author = {Zhang, Chunyi and Yue, Shuwen and Panagiotopoulos, Athanassios Z. and Klein, Michael L. and Wu, Xifan},
  journal = {Phys. Rev. Lett.},
  volume = {131},
  issue = {7},
  pages = {076801},
  numpages = {6},
  year = {2023},
  month = {Aug},
  publisher = {American Physical Society},
  doi = {10.1103/PhysRevLett.131.076801},
  url = {https://link.aps.org/doi/10.1103/PhysRevLett.131.076801}
}

@article{hefter2020dielectric,
  title={Dielectric relaxation spectroscopy: an old-but-new technique for the investigation of electrolyte solutions},
  author={Hefter, Glenn and Buchner, Richard},
  journal={Pure and Applied Chemistry},
  volume={92},
  number={10},
  pages={1595--1609},
  year={2020},
  publisher={De Gruyter}
}

@article{marcus2005electrostriction,
  title={Electrostriction, ion solvation, and solvent release on ion pairing},
  author={Marcus, Yizhak},
  journal={The Journal of Physical Chemistry B},
  volume={109},
  number={39},
  pages={18541--18549},
  year={2005},
  publisher={ACS Publications}
}

@article{galama2013validity,
  title={Validity of the Boltzmann equation to describe Donnan equilibrium at the membrane--solution interface},
  author={Galama, AH and Post, JW and Stuart, MA Cohen and Biesheuvel, PM},
  journal={Journal of membrane science},
  volume={442},
  pages={131--139},
  year={2013},
  publisher={Elsevier}
}

@article{onsager1934surface,
  title={The surface tension of Debye-H{\"u}ckel electrolytes},
  author={Onsager, Lars and Samaras, Nicholas NT},
  journal={The Journal of chemical physics},
  volume={2},
  number={8},
  pages={528--536},
  year={1934},
  publisher={American Institute of Physics}
}

@article{markovich2014surface,
  title={Surface tension of electrolyte solutions: A self-consistent theory},
  author={Markovich, Tomer and Andelman, David and Podgornik, Rudi},
  journal={Europhysics Letters},
  volume={106},
  number={1},
  pages={16002},
  year={2014},
  publisher={IOP Publishing}
}

@article{luo2010simulation,
  title={Simulation of osmotic pressure in concentrated aqueous salt solutions},
  author={Luo, Yun and Roux, Beno{\^\i}t},
  journal={The journal of physical chemistry letters},
  volume={1},
  number={1},
  pages={183--189},
  year={2010},
  publisher={ACS Publications}
}

@article{yoo2018new,
  title={New tricks for old dogs: improving the accuracy of biomolecular force fields by pair-specific corrections to non-bonded interactions},
  author={Yoo, Jejoong and Aksimentiev, Aleksei},
  journal={Physical Chemistry Chemical Physics},
  volume={20},
  number={13},
  pages={8432--8449},
  year={2018},
  publisher={Royal Society of Chemistry}
}

@article{yoo2012improved,
  title={Improved parametrization of Li+, Na+, K+, and Mg2+ ions for all-atom molecular dynamics simulations of nucleic acid systems},
  author={Yoo, Jejoong and Aksimentiev, Aleksei},
  journal={The journal of physical chemistry letters},
  volume={3},
  number={1},
  pages={45--50},
  year={2012},
  publisher={ACS Publications}
}

@article{babu1999new,
  title={A new interpretation of the effective Born radius from simulation and experiment},
  author={Babu, C Satheesan and Lim, Carmay},
  journal={Chemical physics letters},
  volume={310},
  number={1-2},
  pages={225--228},
  year={1999},
  publisher={Elsevier}
}

@article{duignan2013continuum,
  title={A continuum model of solvation energies including electrostatic, dispersion, and cavity contributions},
  author={Duignan, Timothy T and Parsons, Drew F and Ninham, Barry W},
  journal={The Journal of Physical Chemistry B},
  volume={117},
  number={32},
  pages={9421--9429},
  year={2013},
  publisher={ACS Publications}
}

@article{bannon2025influence,
  title={The Influence of Ion Solvation and Association Interactions on Mean Ionic Activity Coefficients in Neutral Polymeric Membranes},
  author={Bannon, Sean M and Fetter, Rachel L and Freger, Viatcheslav and Geise, Geoffrey M},
  journal={Macromolecules},
  volume={58},
  number={22},
  pages={12388--12403},
  year={2025},
  publisher={ACS Publications}
}

@article{pitzer1986thermodynamics,
  title={Thermodynamics of multicomponent, miscible, ionic systems: theory and equations},
  author={Pitzer, Kenneth S and Simonson, John M},
  journal={The journal of physical chemistry},
  volume={90},
  number={13},
  pages={3005--3009},
  year={1986},
  publisher={ACS Publications}
}

@article{marcus2007solvent,
  title={Solvent release upon ion association from entropy data. II},
  author={Marcus, Y},
  journal={The Journal of Physical Chemistry B},
  volume={111},
  number={3},
  pages={572--580},
  year={2007},
  publisher={ACS Publications}
}

@article{shilov2015role,
  title={The role of concentration dependent static permittivity of electrolyte solutions in the Debye--Huckel theory},
  author={Shilov, Ignat Yu and Lyashchenko, Andrey K},
  journal={The Journal of Physical Chemistry B},
  volume={119},
  number={31},
  pages={10087--10095},
  year={2015},
  publisher={ACS Publications}
}

@article{valisko2017activity,
  title={Activity coefficients of individual ions in LaCl3 from the II+ IW theory},
  author={Valisk{\'o}, M{\'o}nika and Boda, Dezs{\H{o}}},
  journal={Molecular Physics},
  volume={115},
  number={9-12},
  pages={1245--1252},
  year={2017},
  publisher={Taylor \& Francis}
}

@article{ben2009beyond,
  title={Beyond standard Poisson--Boltzmann theory: ion-specific interactions in aqueoussolutions},
  author={Ben-Yaakov, Dan and Andelman, David and Harries, Daniel and Podgornik, Rudi},
  journal={Journal of Physics: Condensed Matter},
  volume={21},
  number={42},
  pages={424106},
  year={2009},
  publisher={IOP Publishing}
}

@incollection{bjerrum1968dissociation,
  title={The dissociation of strong electrolytes},
  author={Bjerrum, Niels},
  booktitle={Source Book in Chemistry, 1900--1950},
  pages={181--188},
  year={1968},
  publisher={Harvard University Press}
}

@book{robinson2002electrolyte,
  title={Electrolyte solutions},
  author={Robinson, Robert Anthony and Stokes, Robert Harold},
  year={2002},
  publisher={Courier Corporation}
}

@book{atkins2023atkins,
  title={Atkins' physical chemistry},
  author={Atkins, Peter William and De Paula, Julio and Keeler, James},
  year={2023},
  publisher={Oxford university press}
}

@article{hamer1972osmotic,
  title={Osmotic coefficients and mean activity coefficients of uni-univalent electrolytes in water at 25° C},
  author={Hamer, Walter J and Wu, Yung-Chi},
  journal={Journal of Physical and Chemical Reference Data},
  volume={1},
  number={4},
  pages={1047--1100},
  year={1972},
  publisher={American Institute of Physics for the National Institute of Standards and~…}
}

@article{phillips2005scalable,
  title={Scalable molecular dynamics with NAMD},
  author={Phillips, James C and Braun, Rosemary and Wang, Wei and Gumbart, James and Tajkhorshid, Emad and Villa, Elizabeth and Chipot, Christophe and Skeel, Robert D and Kale, Laxmikant and Schulten, Klaus},
  journal={Journal of computational chemistry},
  volume={26},
  number={16},
  pages={1781--1802},
  year={2005},
  publisher={Wiley Online Library}
}

@article{zheng2017proton,
  title={Proton mobility and thermal conductivities of fuel cell polymer membranes: Molecular dynamics simulation},
  author={Zheng, Chenyang and Geng, Fan and Rao, Zhonghao},
  journal={Computational Materials Science},
  volume={132},
  pages={55--61},
  year={2017},
  publisher={Elsevier}
}

@article{freger2021polyamide,
  title={Polyamide desalination membranes: Formation, structure, and properties},
  author={Freger, Viatcheslav and Ramon, Guy Z},
  journal={Progress in Polymer Science},
  volume={122},
  pages={101451},
  year={2021},
  publisher={Elsevier}
}

@article{werber2016materials,
  title={Materials for next-generation desalination and water purification membranes},
  author={Werber, Jay R and Osuji, Chinedum O and Elimelech, Menachem},
  journal={Nature Reviews Materials},
  volume={1},
  number={5},
  pages={1--15},
  year={2016},
  publisher={Nature Publishing Group}
}

@article{zhou2020intrapore,
  title={Intrapore energy barriers govern ion transport and selectivity of desalination membranes},
  author={Zhou, Xuechen and Wang, Zhangxin and Epsztein, Razi and Zhan, Cheng and Li, Wenlu and Fortner, John D and Pham, Tuan Anh and Kim, Jae-Hong and Elimelech, Menachem},
  journal={Science advances},
  volume={6},
  number={48},
  pages={eabd9045},
  year={2020},
  publisher={American Association for the Advancement of Science}
}

@article{geise2014fundamental,
  title={Fundamental water and salt transport properties of polymeric materials},
  author={Geise, Geoffrey M and Paul, Donald R and Freeman, Benny D},
  journal={Progress in Polymer Science},
  volume={39},
  number={1},
  pages={1--42},
  year={2014},
  publisher={Elsevier}
}

@article{epsztein2020towards,
  title={Towards single-species selectivity of membranes with subnanometre pores},
  author={Epsztein, Razi and DuChanois, Ryan M and Ritt, Cody L and Noy, Aleksandr and Elimelech, Menachem},
  journal={Nature Nanotechnology},
  volume={15},
  number={6},
  pages={426--436},
  year={2020},
  publisher={Nature Publishing Group UK London}
}

@article{fu2025biomimetic,
  title={Biomimetic ion channels with subnanometer sizes for ion sieving: a mini-review},
  author={Fu, Qianqian and Ma, Zhaoyu and Gao, Jun},
  journal={Nanoscale},
  year={2025},
  publisher={Royal Society of Chemistry}
}

@article{li2025constructing,
  title={Constructing new-generation ion exchange membranes under confinement regime},
  author={Li, Xingya and Zuo, Peipei and Ge, Xiaolin and Yang, Zhengjin and Xu, Tongwen},
  journal={National Science Review},
  volume={12},
  number={2},
  pages={nwae439},
  year={2025},
  publisher={Oxford University Press}
}

@article{li2023designing,
  title={Designing artificial ion channels with strict K+/Na+ selectivity toward next-generation electric-eel-mimetic ionic power generation},
  author={Li, Jipeng and Du, Linhan and Kong, Xian and Wu, Jianzhong and Lu, Diannan and Jiang, Lei and Guo, Wei},
  journal={National Science Review},
  volume={10},
  number={12},
  pages={nwad260},
  year={2023},
  publisher={Oxford University Press}
}

@book{strathmann2004ion,
  title={Ion-exchange membrane separation processes},
  author={Strathmann, Heiner},
  volume={9},
  year={2004},
  publisher={Elsevier}
}

@article{yaroshchuk2019modelling,
  title={Modelling nanofiltration of electrolyte solutions},
  author={Yaroshchuk, Andriy and Bruening, Merlin L and Zholkovskiy, Emiliy},
  journal={Advances in colloid and interface science},
  volume={268},
  pages={39--63},
  year={2019},
  publisher={Elsevier}
}

@article{kingsbury2020comparison,
  title={Comparison of water and salt transport properties of ion exchange, reverse osmosis, and nanofiltration membranes for desalination and energy applications},
  author={Kingsbury, RS and Wang, J and Coronell, O},
  journal={Journal of Membrane Science},
  volume={604},
  pages={117998},
  year={2020},
  publisher={Elsevier}
}

@article{biesheuvel2022tutorial,
  title={Tutorial review of reverse osmosis and electrodialysis},
  author={Biesheuvel, PM and Porada, S and Elimelech, M and Dykstra, JE},
  journal={Journal of Membrane Science},
  volume={647},
  pages={120221},
  year={2022},
  publisher={Elsevier}
}

@book{barthel1998physical,
  title={Physical chemistry of electrolyte solutions: modern aspects},
  author={Barthel, Josef and Krienke, Hartmut and Kunz, Werner and Kunz, W},
  volume={5},
  year={1998},
  publisher={Springer Science \& Business Media}
}

@article{grosberg2002colloquium,
  title={Colloquium: The physics of charge inversion in chemical and biological systems},
  author={Grosberg, A Yu and Nguyen, TT and Shklovskii, BI},
  journal={Reviews of modern physics},
  volume={74},
  number={2},
  pages={329},
  year={2002},
  publisher={APS}
}

@article{freger2023dielectric,
  title={Dielectric exclusion, an {\'e}minence grise},
  author={Freger, Viatcheslav},
  journal={advances in colloid and interface science},
  volume={319},
  pages={102972},
  year={2023},
  publisher={Elsevier}
}

@book{hill2012introduction,
  title={An introduction to statistical thermodynamics},
  author={Hill, Terrell L},
  year={2012},
  publisher={Courier Corporation}
}

@article{freger2025ion,
  title={Ion uptake and pairing in membranes: The pore model},
  author={Freger, Viatcheslav},
  journal={Journal of Membrane Science},
  volume={722},
  pages={123795},
  year={2025},
  publisher={Elsevier}
}

@article{freger2020ion,
  title={Ion partitioning and permeation in charged low-T* membranes},
  author={Freger, Viatcheslav},
  journal={Advances in Colloid and Interface Science},
  volume={277},
  pages={102107},
  year={2020},
  publisher={Elsevier}
}

@article{oren2024analyzing,
  title={Analyzing ion uptake in ion-exchange membranes using ion association model},
  author={Oren, Yaeli S and Nir, Oded and Freger, Viatcheslav},
  journal={Journal of Membrane Science},
  volume={690},
  pages={122202},
  year={2024},
  publisher={Elsevier}
}

@article{colbin2025ion,
  title={Ion-Pairing: A Bygone Treatment of Electrolyte Solutions?},
  author={Colbin, Lars Olow Simon and Shao, Yunqi and Younesi, Reza},
  journal={Batteries \& Supercaps},
  volume={8},
  number={1},
  pages={e202400160},
  year={2025},
  publisher={Wiley Online Library}
}

@article{roy2017marcus,
  title={Marcus theory of ion-pairing},
  author={Roy, Santanu and Baer, Marcel D and Mundy, Christopher J and Schenter, Gregory K},
  journal={Journal of chemical theory and computation},
  volume={13},
  number={8},
  pages={3470--3477},
  year={2017},
  publisher={ACS Publications}
}

@article{fennell2009ion,
  title={Ion pairing in molecular simulations of aqueous alkali halide solutions},
  author={Fennell, Christopher J and Bizjak, Alan and Vlachy, Vojko and Dill, Ken A},
  journal={The Journal of Physical Chemistry B},
  volume={113},
  number={19},
  pages={6782--6791},
  year={2009},
  publisher={ACS Publications}
}

@article{marcus2006ion,
  title={Ion pairing},
  author={Marcus, Yizhak and Hefter, Glenn},
  journal={Chemical reviews},
  volume={106},
  number={11},
  pages={4585--4621},
  year={2006},
  publisher={ACS Publications}
}

@article{gierst1966ion,
  title={Ion pairing mechanisms in electrode processes},
  author={Gierst, Lucien and Vandenberghen, Lucienne and Nicolas, Edgard and Fraboni, Alain},
  journal={Journal of The Electrochemical Society},
  volume={113},
  number={10},
  pages={1025},
  year={1966},
  publisher={IOP Publishing}
}

@article{saveant2001effect,
  title={Effect of ion pairing on the mechanism and rate of electron transfer. Electrochemical aspects},
  author={Sav{\'e}ant, Jean-Michel},
  journal={The Journal of Physical Chemistry B},
  volume={105},
  number={37},
  pages={8995--9001},
  year={2001},
  publisher={ACS Publications}
}

@article{marcus1998ion,
  title={Ion pairing and electron transfer},
  author={Marcus, RA},
  journal={The Journal of Physical Chemistry B},
  volume={102},
  number={49},
  pages={10071--10077},
  year={1998},
  publisher={ACS Publications}
}

@article{hefter2006spectroscopy,
  title={When spectroscopy fails: The measurement of ion pairing},
  author={Hefter, Glenn},
  journal={Pure and applied chemistry},
  volume={78},
  number={8},
  pages={1571--1586},
  year={2006},
  publisher={International Union of Pure and Applied Chemistry}
}

@article{saveant2008evidence,
  title={Evidence for concerted pathways in ion-pairing coupled electron transfers},
  author={Sav{\'e}ant, Jean-Michel},
  journal={Journal of the American Chemical Society},
  volume={130},
  number={14},
  pages={4732--4741},
  year={2008},
  publisher={ACS Publications}
}

@article{mason2019molecular,
  title={Molecular dynamics and neutron scattering studies of potassium chloride in aqueous solution},
  author={Mason, Philip E and Tavagnacco, Letizia and Saboungi, Marie-Louise and Hansen, Thomas and Fischer, Henry E and Neilson, George W and Ichiye, Toshiko and Brady, John W},
  journal={The Journal of Physical Chemistry B},
  volume={123},
  number={50},
  pages={10807--10813},
  year={2019},
  publisher={ACS Publications}
}

@article{kohagen2016accounting,
  title={Accounting for electronic polarization effects in aqueous sodium chloride via molecular dynamics aided by neutron scattering},
  author={Kohagen, Miriam and Mason, Philip E and Jungwirth, Pavel},
  journal={The Journal of Physical Chemistry B},
  volume={120},
  number={8},
  pages={1454--1460},
  year={2016},
  publisher={ACS Publications}
}

@article{harsanyi2012neutron,
  title={Neutron and X-ray diffraction measurements on highly concentrated aqueous LiCl solutions},
  author={Hars{\'a}nyi, Ildik{\'o} and Temleitner, L{\'a}szl{\'o} and Beuneu, Brigitte and Pusztai, L{\'a}szl{\'o}},
  journal={Journal of Molecular Liquids},
  volume={165},
  pages={94--100},
  year={2012},
  publisher={Elsevier}
}

@article{searle1995application,
  title={Application of a generalised enthalpy--entropy relationship to binding co-operativity and weak associations in solution},
  author={Searle, Mark S and Westwell, Martin S and Williams, Dudley H},
  journal={Journal of the Chemical Society, Perkin Transactions 2},
  volume={24},
  number={1},
  pages={141--151},
  year={1995},
  publisher={The Royal Society of Chemistry}
}

@article{bouazizi2008structural,
  title={Structural investigations of high concentrated aqueous LiCl solutions: X-ray scattering and MD simulations approach},
  author={Bouazizi, Salah and Nasr, Salah},
  journal={Journal of Molecular Structure},
  volume={875},
  number={1-3},
  pages={121--129},
  year={2008},
  publisher={Elsevier}
}

@article{bouazizi2006local,
  title={Local order in aqueous NaCl solutions and pure water: X-ray scattering and molecular dynamics simulations study},
  author={Bouazizi, Salah and Nasr, Salah and Ja{\^\i}dane, Nejmeddine and Bellissent-Funel, Marie-Claire},
  journal={The Journal of Physical Chemistry B},
  volume={110},
  number={46},
  pages={23515--23523},
  year={2006},
  publisher={ACS Publications}
}

@article{abascal2005general,
  title={A general purpose model for the condensed phases of water: TIP4P/2005},
  author={Abascal, Jose LF and Vega, Carlos},
  journal={The Journal of chemical physics},
  volume={123},
  number={23},
  year={2005},
  publisher={AIP Publishing}
}

@article{fox2018molecular,
  title={The molecular origin of enthalpy/entropy compensation in biomolecular recognition},
  author={Fox, Jerome M and Zhao, Mengxia and Fink, Michael J and Kang, Kyungtae and Whitesides, George M},
  journal={Annual Review of Biophysics},
  volume={47},
  number={1},
  pages={223--250},
  year={2018},
  publisher={Annual Reviews}
}

@article{ryde2014fundamental,
  title={A fundamental view of enthalpy--entropy compensation},
  author={Ryde, Ulf},
  journal={MedChemComm},
  volume={5},
  number={9},
  pages={1324--1336},
  year={2014},
  publisher={The Royal Society of Chemistry}
}

\end{document}


\setcounter{secnumdepth}{1}

\renewcommand{\theequation}{S\arabic{equation}}
\renewcommand{\thefigure}{S\arabic{figure}}
\renewcommand{\thesection}{S\arabic{section}}

\setcounter{equation}{0}
\setcounter{figure}{0}
\setcounter{section}{0}

\preprint{AIP/123-QED}

\title {Supplementary Material\\Ion-Pairing Enhancement under Osmotic Stress: Disentangling the Effects of Ion and Water Activities}


\author{Jay Prakash Singh}
\affiliation{%
Wolfson Department of Chemical Engineering, Technion Israel Institute of Technology, Haifa, Israel%
}

\author{Viatcheslav Freger$^{*}$}
\email{vfreger@technion.ac.il}
\affiliation{%
Wolfson Department of Chemical Engineering, Technion Israel Institute of Technology, Haifa, Israel%
}
\affiliation{%
Grand Technion Energy Program, Technion Israel Institute of Technology, Haifa, Israel%
}
\affiliation{%
Grand Water Research Institute, Technion Israel Institute of Technology, Haifa, Israel%
}

\date{\today}

\maketitle

\begin{figure*}
    \centering
    \includegraphics[width=0.9\linewidth]{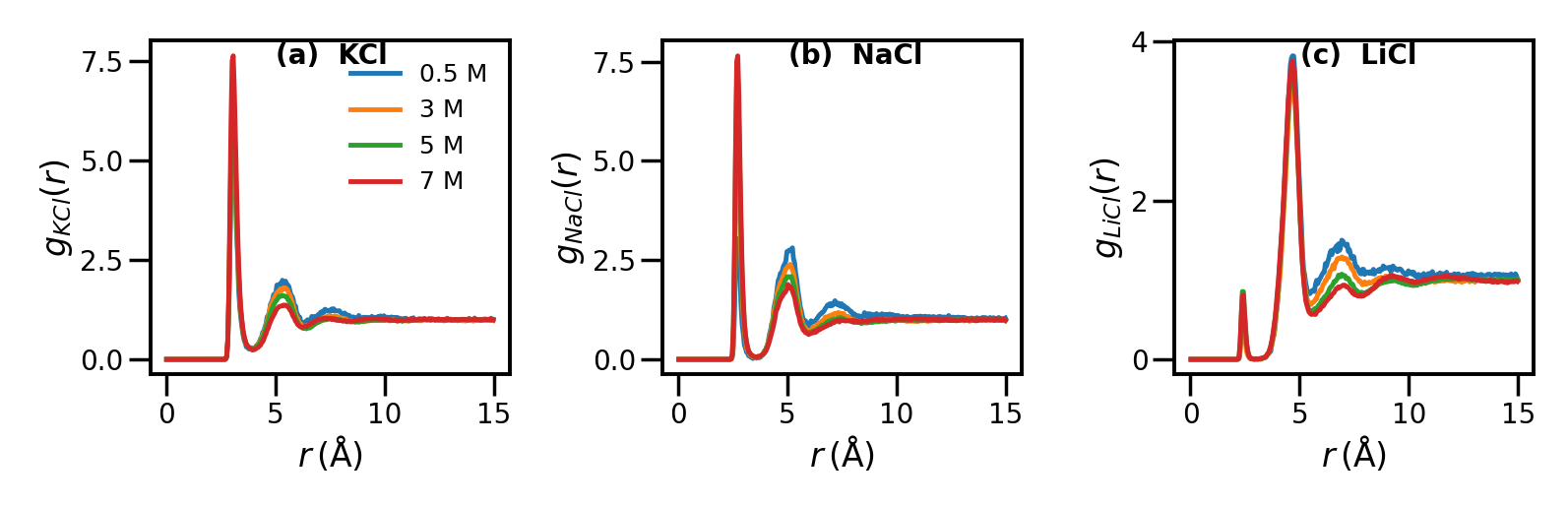}
   \caption{
Radial distribution functions (RDFs) for monovalent salts in the volume containing the salt at different concentrations within this volume. 
Panels (a), (b), and (c) correspond to KCl, NaCl, and LiCl, respectively. 
For each salt, RDFs are shown for the concentrations indicated in panel (a). 
The concentrations were increased by adjusting the width of the zero-potenital region of confining potential for the same amount of salt. Distinct peaks of RDF identify the different ion-pairing states, namely, contact ion pairs (CIP), solvent-shared ion pairs (SIP), and solvent-separated ion pairs (SSIP), and the minima separate between the states.
}
    \label{fig:radsm}
\end{figure*}

\section{ Computed RDF and Association Constants}
\vspace{-1em} 
Cation-anion radial distribution functions (RDFs) $g_{\mathrm{M{-}A}}(r)$ computed for monovalent salts (KCl, NaCl, and LiCl) for several representative concentrations within the solution volume, defined by the shape of the ion-confining potential, are shown in Fig.~\ref{fig:radsm}. 
The computed RDFs for all three salts are consistent with available neutron and X-ray scattering data, reproducing the experimentally observed positions of the main maxima and minima in the cation-anion distance distributions that define the characteristic ion-pairing states, thereby validating the force-field description of ion-ion correlations used in the present simulations. The RDF also reproduce reasonably relative magnitude of the peaks, subject to inherent noise and uncertainties of separating the cation-anion RDF form intense scattering background of water-water and ion-water correlations~\cite{mason2019molecular,kohagen2016accounting,harsanyi2012neutron,bouazizi2008structural}.  
Distinct peaks in the RDFs, corresponding to progressively increasing cation-anion separations identify the different ion-pairing states, namely contact ion pairs (CIP), solvent-shared ion pairs (SIP), and solvent-separated ion pairs (SSIP). Based on these RDFs, the corresponding concentration-based association constants $K_{\mathrm{CIP}}$, $K_{\mathrm{SIP}}$, and $K_{\mathrm{SSIP}}$ were obtained by integrating the RDFs over appropriate distance ranges defined by the positions of successive minima, $r_{\min}$ and $r_{\max}$, as 
$K = 4\pi \int_{r_{\min}}^{r_{\max}} g_{\mathrm{M{-}A}}(r)\, r^2\, dr$. 
Specifically, CIP are defined from zero separation to the first minimum, SIP from the first to the second minimum, and SSIP from the second to the third minimum (for LiCl only). The integration yields association constants in units of volume ($\text{\AA}^3$).

They were first converted to $\mathrm{L\,mol^{-1}}$ and subsequently transformed into mole-fraction-based association constants by multiplying by the total molar concentration of relevant species (here, total salt = total cation = total anion and water) within the confined volume. For NaCl and KCl, CIP and SIP populations are dominant, whereas SSIP are weakly pronounced and thus were lumped into the free-ion state. In contrast, LiCl exhibits a substantially populated SSIP state, which is therefore explicitly included in the analysis.

\section{Computed water activities: comparison with experiment}
\vspace{-1em} 
To examine the agreement of the force-field parameters used in simulations with experiments, we compared the relation between salt concentration and water activity obtaiend in simulations with experimental data for the three salts\cite{hamer1972osmotic}. We also tested agreement with the corresponding BET isotherm (see Eq. 8 in the main text). The experimental water activities \(a_w\) were computed directly from the molality \(m\) and reported osmotic coefficient \(\phi\) of the solutions using the relation $\ln a_w = -2\, M_w \, \phi \, m,$
where \(M_w = 0.018015~\mathrm{kg\,mol^{-1}}\) is the molar mass of water \cite{hamer1972osmotic}. The total number of water molecules in solution per salt , \(q = N_w/N_s\), was then calculated from the molality, as $q = 1/m M_w$.  Thus obtained $q$ could be fitted to the BET equation.

\begin{figure}
    \centering    \includegraphics[width=1.0\linewidth]{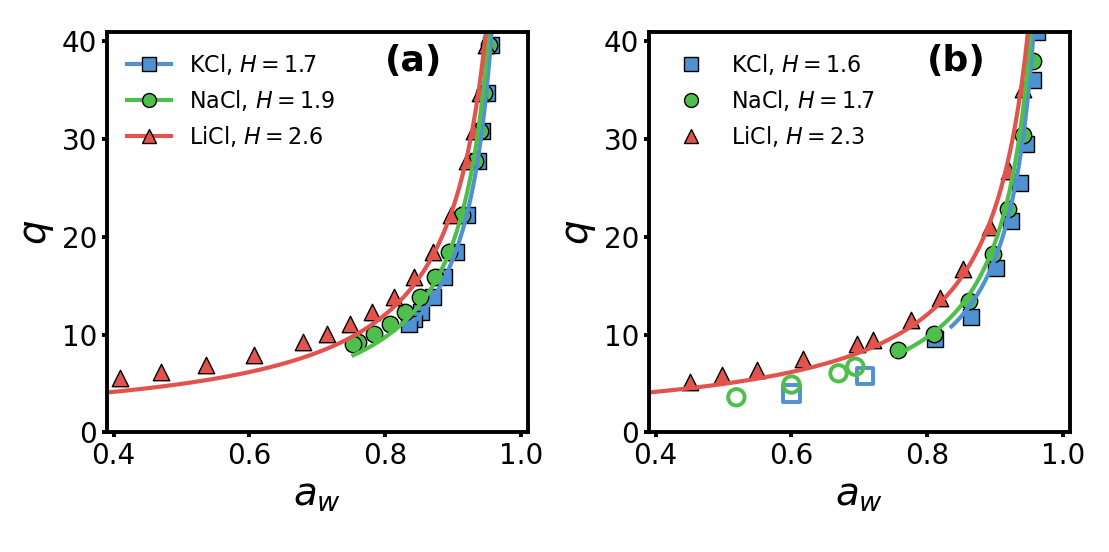}
    \caption{
Comparison of computed and experimental dependence of water activity on solution concentration for KCl, NaCl, and LiCl, based on reported osmotic coefficients \cite{hamer1972osmotic} and plotted as the water-to-salt molar ratio $q$ versus $a_w$ and corresponding BET model fits. 
Panel (a) shows experimental data as symbols and the corresponding BET isotherm fits as lines. 
Panel (b) compares the experimental data, shown as solid line interpolating between the data points, and simulated results shown as symbols. 
The legends indicate hydration number $H$, obtained by fitting the BET isotherm (see Eq. 8 in the main text), to experimental (a) and simulated (b) data. Open symbols represent data points above the solubility limits of the respective salts.  
}
    \label{fig:bet}
\end{figure}

First, Fig. \ref{fig:bet}(a) compares the experimental data as $q$ versus $a_w$ (symbols) and corresponding BET isotherm fits (lines), which match the data well. Subsequently, Fig. \ref{fig:bet}(b) compares the experimental data, displayed as solid lines interpolating between experimental data points, with the simulation results (symbols). Some deviations may be noted, and the BET fits to the experimental (Fig. \ref{fig:bet}(a)) and simulated (fits not shown) data yield slightly different hydration numbers \(H\), indicated in the legends. Nevertheless, the simulated and experimental trends and \(H\) values match reasonably closely.

We note that, for very soluble LiCl, the entire simulated range remains within the solubility limit and can be compared with experimental values. However, for KCl and NaCl, the physical concentration range is bounded by their solubility limits. To see how their trends extrapolate, we also explored for KCl and NaCl and displayed in Fig. \ref{fig:bet}(b) the non-physical concentrations that artificially exceed the experimental solubility limits. Note that, since MD force fields were not adjusted and were not meant to accurate reproduce the thermodynamic properties of the solid salt phase, they were not supposed to reproduce the solubility limits and thus MD could simulate the non-physical range beyond solubility.  It is seen that in this non-physical range, the extrapolated trends of KCl and NaCl behave very similar to that of LiCl.

\section{Computed ion activities: comparison with experiment}
\begin{figure}
    \centering    \includegraphics[width=0.8\linewidth]{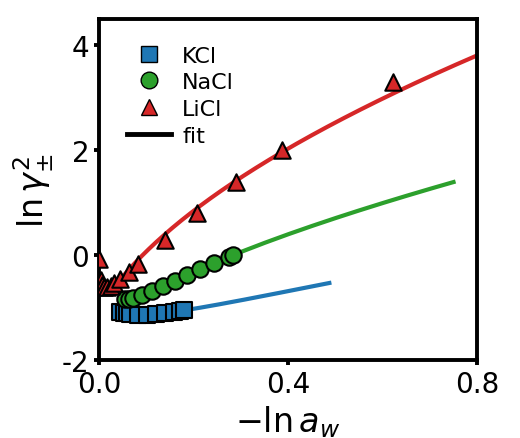}
    \caption{Dependence of the ion activity term $\ln \gamma^2_{\pm}$ on the negative logarithm of water activity $-\ln a_w$. Experimental data (symbols) were taken from Ref.\cite{hamer1972osmotic}. Solid lines represent fits using Eq. 8 in the main text. The fitting was performed only over the regions used in the analysis and constrained by the availability of experimental data. For KCl and NaCl only a few data points are available; nevertheless, the fitted relation appears to extrapolate reasonably well to the higher $-\ln a_w$ region used in the calculations.}

    \label{fig:ionc}
\end{figure}

\vspace{-1em} 
As further test of the agreement between the force-field parameters used in the simulations and experimental observations, we compared the ion activity coefficients, computed as described in the next section using the fitted BET parameters for simulated water activity vs. salt concentration, with available experimental data for the three salts, as shown in Fig. \ref{fig:ionc}. The tested data rage of salt concentrations is above 1~M for all three salts, corresponding to $-\ln a_w \ge 0.06$. Lower concentrations were not included here, as the computed osmotic pressure was excessively noisy for this range. Notably, this dilute regime also shows a different trend, approximately described by the Debye-Hückel theory, that can not be reproduced by the BET isotherm. Since the present focus was on the region where the osmotic pressure is relatively large, excluding this dilute region was not an issue. 

Fig. \ref{fig:ionc} shows that, in the physical range below solubility limit, the BET fits for KCl and NaCl agree well with the experimental values, with deviations of only about 1-2 \%. These small differences might also reflect uncertainties or scatter of the experimental measurements. For LiCl, the points derived from the simulated osmotic data and BET fits deviates more significantly, by about 5 \% in the concentration range 1-6~M, and the deviation increases to approximately 10 \% at highest concentration, due to poorer fit of BET in this range, which was still considered reasonable for the present purpose.

\section{Separation of the ion and water activity contributions}
\vspace{-1em} 
The dependence of the concentration-based association constant $K(a_w)$, as defined in Eq.~(4) of the main text, on water activity reflects two distinct physical contributions. The first is indirect and arises from the activity coefficients of ions and ion-pairs, reflecting the changes in ion-ion and ion-water interactions and correlations, varying with the osmotic pressure. The second is directly related to water activity and corresponds to an the effect of water release accompanying ion-pair formation. 

The hydration function $q(a_w) = (1-x_s)/x_s$, defined as the total number of water molecules in solution per salt "molecule", was described by the BET-type multilayer adsorption model (Eq. 8 in the main text). Using this model, the relation $a_s=(\gamma_\pm x_s)^2$, and the Gibbs-Duhem equation, the variation of the mean ionic activity term can be related to $q(a_w)$ as;

\begin{equation}
-\frac{d\ln\gamma^2_\pm}{d\ln a_w}
=
\Big[q+2\frac{d\ln x_s}{d\ln a_w}\Big].
\end{equation}

Considering next the general differential relation, eq. 4 in the main paper, governing the activity dependence of the experimentally measured association constant, we define the ``corrected" association constant $\widetilde{K}$, in which the effect of salt non-ideality (ion-activity coefficient $\gamma^2_\pm$) is removed. The slope of the variation of thus obtained $\widetilde{K}$ with the osmotic pressure $-\ln a_w$ directly yields the parameter of interest $\Delta n^*$ that combines the effects of water shedding and ion-pair non-ideality as follows 
\begin{equation}
\begin{split}
\label{eq:ktilde}
-\frac{d\ln \widetilde{K}}{d\ln a_w}
=
-\frac{d\ln K}{d\ln a_w}
+
\frac{d\ln\gamma^2_\pm}{d\ln a_w}
=
 \Delta n^*,
 \end{split}
\end{equation}
where $\Delta n^*$ is the effective number of water molecules expressing the combined effects of water activity $a_w$ through actual shedding of water molecules ($\Delta n$) and varying non-ideality of ion-pairs embedded in  $\gamma_p$ in Eqs. 3 and 4. This expression may be recast in integral form to express the ion-activity ``correction" to $K$ as  

\begin{equation}
\begin{split}
 \ln \widetilde{K} = \ln K-\ln \gamma_{\pm}^2
= 
\ln K-\int_{1}^{a_w} d\ln \gamma_{\pm}^2 \\
=
\ln K-\int_{1}^{a_w}
\left[q
+
2\frac{d\ln x_s}{d\ln a_w}
\right]
d\ln a_w .
\label{eq:int_corr}   
\end{split}
\end{equation}
Here the integration is performed between the states of the pure solvent ($a_w=1$ and $x_s=0$) as the reference state, for which we set $\widetilde{K}=K$ and $\gamma_{\pm}=1$, to the desired water activity $a_w$.

\begin{figure}
    \centering
    \includegraphics[width=0.75\linewidth]{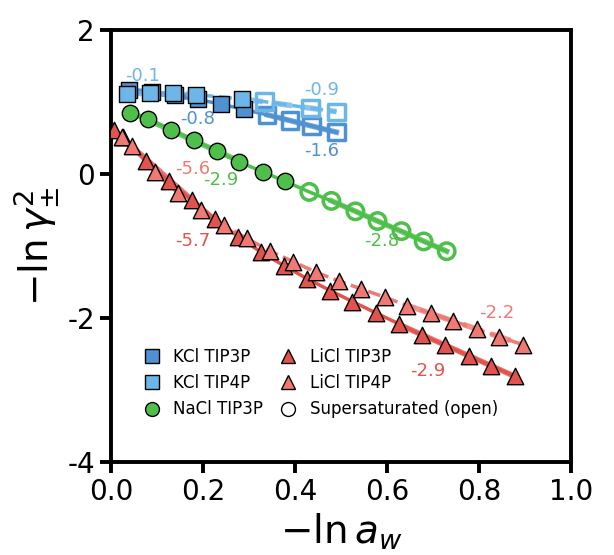}
    
 \caption{
The dependence of the mean ionic activity correction to the concentration-based pairing constant, $-\ln \gamma_{\pm}^{2}$, calculated using Eq.~\ref{eq:master_aw}, on water activity for KCl, NaCl, and LiCl, shown as $-\ln \gamma_{\pm}^{2}$ versus $-\ln a_w$. Open symbols denote data points above the solubility limit of the corresponding salt, while filled symbols represent data below the solubility limit. The solid lines indicate linear fits to the data in the respective regimes, with the corresponding slopes indicated in the plot. } 

  \label{fig:ioncor}
\end{figure}

To perform the last integral, the simulated dependence of $q(a_w)$ fitted to the BET isotherm, Eq. 8, along with the relation $x_s=1/(q+1)$ might be used, yielding a tractable analytical expression for the anti-derivative. In this manner, the ion-activity correction is linked to and deduced from the computed osmotic data. Unfortunately, the known flaw of the BET model is that it extrapolates nonphysically to the infinite dilution \cite{hill2012introduction}, thereby the resulting integrated expression diverges at $a_w=1$ . However, given the present MD results for osmotic pressure (hence $a_w$) reasonably agree with experimental data within the simulated range, we use the BET fit only for interpolating between simulated data points, i.e., integration within the simulated range of $a_w \leq a_w^{(1)}$, where $a_w^{(1)}$ is the water activity of the most dilute simulated solution. The ion-activity correction is then computed as follows
\begin{equation}
\begin{split}
\ln \gamma_{\pm}^2(a_w)
&= 
\int_{1}^{a_w^{(1)}} d\ln \gamma^2_{\pm}
+
\int_{a_w^{(1)}}^{a_w} d\ln \gamma^2_{\pm} \\
&= 
\ln \gamma^2_{\pm}(a_w^{(1)})
-
\Bigg[ H\ln \frac{(C-1)a_w+1}{1-a_w}
-
2\ln (q+1) \Bigg]_{a_w^{(1)}}^{a_w}
\label{eq:master_aw}
\end{split}
\end{equation}
Here, integration in the range $a_w^{(1)} < a_w \leq 1$ yields $\ln \gamma_{\pm}(a_w^{(1)})$, for which we use the known experimental value  \cite{hamer1972osmotic}. Notably, this constant value has no effect on the differential relation, Eq.~\ref{eq:ktilde}, hence on $\Delta n^*$. However, due to non-monotonic trend of $\ln \gamma_{\pm}(a_w)$ at lower concentrations, well observed in Fig.~\ref{fig:ionc}, the trends in Fig.~\ref{fig:ioncor} do not extrapolate to $\ln \gamma_{\pm}=0$  at zero osmotic pressure.

Thus computed ion-activity correction $-\ln \gamma_{\pm}^2$  is plotted in Fig. \ref{fig:ioncor}. It is notable that the slope tends to become less negative with the increasing osmotic pressure, which might partly reflect the increasing error of BET fits for LiCl. However, a similar but less pronounced trend is also observed for KCl and NaCl, for which the BET fits are fairly accurate. The resulting corrected association constants, as $\ln \widetilde{K} = \ln K -\ln \gamma_{\pm}^2$ for different ion-pairing states (CIP and SIP as well as SSIP  for LiCl), are plotted vs. $-\ln a_w$ in Fig. 3 (a-b) shown in the main text.

\appendix
\renewcommand\refname{ references}

\bibliography{aipsamp}